\documentclass[manuscript,screen,nonacm]{acmart}

\renewcommand\footnotetextcopyrightpermission[1]{}

\AtBeginDocument{%
  }

\setcopyright{acmlicensed}
\copyrightyear{2018}
\acmYear{2018}
\acmDOI{XXXXXXX.XXXXXXX}
\acmConference[Conference acronym 'XX]{Make sure to enter the correct
  conference title from your rights confirmation email}{June 03--05,
  2018}{Woodstock, NY}
\acmISBN{978-1-4503-XXXX-X/2018/06}

\usepackage[edges]{forest}
\usepackage{adjustbox}
\usepackage{graphicx}  
\usepackage{float} 
\usepackage{amsmath}

\usepackage{amsfonts}

\definecolor{hidden-draw}{RGB}{20,68,106}
\definecolor{hidden-pink}{RGB}{255,245,247}

\usepackage{booktabs}
\usepackage{tabularx}
\usepackage{enumitem}
\usepackage{subcaption}
\usepackage{array}
\usepackage{enumitem}

\newcolumntype{L}[1]{>{\raggedright\arraybackslash}p{#1}}
\newcolumntype{C}[1]{>{\centering\arraybackslash}p{#1}}
\newcolumntype{Y}{>{\raggedright\arraybackslash}X}
\newcolumntype{M}[1]{>{\centering\arraybackslash}m{#1}}

\usepackage{forest}
\usepackage{adjustbox}
\usepackage{tikz}
\usetikzlibrary{shadows.blur}

\begin{document}

\title{Query Expansion in the Age of Pre-trained and Large Language Models: A Comprehensive Survey}

\author{Minghan Li}
\authornote{Corresponding author.}
\affiliation{%
  \institution{School of Computer Science and Technology, Soochow University}
  \city{Suzhou}
  \state{Jiangsu}
  \postcode{215031}
  \country{China}
}
\email{mhli@suda.edu.cn}

\author{Xinxuan Lv}
\affiliation{%
  \institution{School of Computer Science and Technology, Soochow University}
  \city{Suzhou}
  \state{Jiangsu}
  \postcode{215031}
  \country{China}
}
\email{xxlv@stu.suda.edu.cn}

\author{Junjie Zou}
\affiliation{%
  \institution{School of Computer Science and Technology, Soochow University}
  \city{Suzhou}
  \state{Jiangsu}
  \postcode{215031}
  \country{China}
}
\email{jjzou1@stu.suda.edu.cn}

\author{Tongna Chen}
\affiliation{%
  \institution{School of Computer Science and Technology, Soochow University}
  \city{Suzhou}
  \state{Jiangsu}
  \postcode{215031}
  \country{China}
}
\email{tnchentnchen@stu.suda.edu.cn}

\author{Chao Zhang}
\affiliation{%
  \institution{School of Computer Science and Technology, Soochow University}
  \city{Suzhou}
  \state{Jiangsu}
  \postcode{215031}
  \country{China}
}
\email{czhang1@stu.suda.edu.cn}

\author{Suchao An}
\affiliation{%
  \institution{School of Computer Science and Technology, Soochow University}
  \city{Suzhou}
  \state{Jiangsu}
  \postcode{215031}
  \country{China}
}
\email{scan@stu.suda.edu.cn}

\author{Ercong Nie}
\affiliation{%
  \institution{Center for Information and Language Processing (CIS), LMU Munich}
  \city{Munich}
  \postcode{80538}
  \country{Germany}
}
\affiliation{%
  \institution{Munich Center for Machine Learning (MCML)}
  \city{Munich}
  \postcode{80538}
  \country{Germany}
}
\email{nie@cis.lmu.de}

\author{Guodong Zhou}
\affiliation{%
  \institution{School of Computer Science and Technology, Soochow University}
  \city{Suzhou}
  \state{Jiangsu}
  \postcode{215031}
  \country{China}
}
\email{gdzhou@suda.edu.cn}

\renewcommand{\shortauthors}{Li et al.}

\begin{abstract}

Modern information retrieval must reconcile short, ambiguous queries with increasingly diverse and dynamic corpora. Query expansion (QE) remains a core technique for mitigating vocabulary mismatch, but its design space has been reshaped by pre-trained and large language models (PLMs/LLMs). This survey reviews QE methods in the PLM/LLM era and provides a unified view of the emerging landscape. We first summarize how different model families enable new expansion behaviors, including stronger contextualization, more controllable generation, and instruction-following. We then organize recent techniques along four complementary design dimensions: where expansion is injected in the pipeline, how it is grounded and interacts with corpus evidence, how it is learned or aligned, and how structured knowledge such as knowledge graphs is incorporated. Beyond taxonomy, we synthesize application patterns and deployment considerations across representative retrieval settings, highlighting practical trade-offs among effectiveness, controllability, grounding quality, and operating cost. Finally, we outline open challenges and future directions toward more reliable, safe, efficient, and continually adaptive QE under real-world constraints\footnote{Resources are available at \url{https://github.com/lmh0921/QueryExpansion-PLM-LLM-Survey-paperList}}.
\end{abstract}

\ccsdesc[500]{Information systems~Information retrieval query processing}

\keywords{Query Expansion, Information Retrieval, Large Language Models}

\maketitle

\section{Introduction}
\label{sec:intro}

Modern IR systems must interpret increasingly short, context-poor queries issued via mobile, voice, and conversational interfaces, while the available content grows in scale, diversity, and linguistic complexity. The resulting vocabulary mismatch---users express intents with general or ambiguous terms whereas relevant documents use domain-specific terminology, paraphrases, or emerging jargon---is a long-standing obstacle. Context is often omitted, further hindering accurate intent inference. QE addresses these challenges by enriching the initial query with semantically related, contextually appropriate material so as to increase overlap with relevant documents.

Classical QE spans pseudo-relevance feedback (PRF), thesaurus/ontology expansion, translation-based models, and log-driven reformulation~\citep{rocchio1971relevance,voorhees1994query,lavrenko2017relevance}. However, these techniques largely exploit lexical co-occurrence or static resources, making them brittle for short, ambiguous, and long-tail queries and prone to drift and limited coverage~\citep{voorhees1994query}. 

Recent advances in PLMs and LLMs---e.g., BERT/RoBERTa-style encoders for context-sensitive representations~\citep{nogueira2019passage}, encoder--decoders such as T5/BART for controlled generation, and decoder-only LLMs (e.g., GPT-3/4, PaLM, LLaMA-family) for zero-/few-shot reasoning---have opened a broader design space for QE. These models support implicit expansion in the embedding space (e.g., PRF with dense vectors), selection-based term filtering with contextual encoders, and generative expansion via pseudo-documents or structured rationales. For instance, Query2Doc \citep{wang-etal-2023-query2doc} synthesizes pseudo-documents and yields 3--15\% gains on MS MARCO~\cite{DBLP:conf/nips/NguyenRSGTMD16} and TREC DL \citep{craswell2025overview}. Instruction tuning and in-context learning further improve format adherence and zero-shot robustness \citep{zhang2024instruction}. Open challenges persist: factual grounding, domain transfer under data sparsity, precision control for generative outputs, and cost/latency constraints at deployment.

Query expansion, as a classic problem in the field of information retrieval, has been systematically reviewed by numerous scholars. Early survey works, such as the study by \citet{bhogal2007review} provided a comprehensive overview of semantic QE; \citet{sartori2009comparison} focused on ontology-driven QE, pointing out that ontologies can help obtain higher-quality expansion terms;~\citet{lei2011survey} summarize the QE based on local analysis; \citet{carpineto2012survey} offered a brief summary of automatic QE, covering techniques based on linguistics, corpora, the query itself, retrieval logs, and Web data, and discussed related key issues; \citet{rivas2014study} confirmed that combining multiple classical techniques can effectively improve retrieval performance; Subsequently \citet{raza2019survey} systematically discussed statistical QE including document analysis, retrieval and browsing log analysis, and Web knowledge analysis, while \citet{azad2019query} reviewed the core techniques, data sources, weighting and ranking methods, user involvement, and application scenarios of QE. However, these existing surveys primarily focus on traditional technical frameworks. In recent years, deep learning, represented by PLM and LLM, has achieved breakthroughs, bringing new opportunities and challenges to QE with their powerful semantic understanding and generation capabilities. Although new QE methods based on modern language models are emerging,
to the best of our knowledge, there is still no dedicated, IR-focused, deployment-oriented survey centered on PLM/LLM-based QE.
Therefore, this paper aims to fill this gap by systematically reviewing, classifying, and prospecting query expansion techniques that integrate modern PLM/LLM, providing a clear landscape and valuable reference for subsequent research in this field.

This survey synthesizes the landscape of QE in the PLM/LLM era with three unifying perspectives:

\vspace{0.5em} 

\begin{itemize}
    \setlength{\leftmargin}{1.5em}
    \setlength{\labelwidth}{0.5em}
    \setlength{\labelsep}{0.5em}
    \item[\textemdash] \textit{Model taxonomy.} We categorize PLMs/LLMs relevant to QE and articulate their characteristic affordances for QE.
    \item[\textemdash] \textit{Paradigms.} We trace the evolution from symbolic/statistical QE to implicit/embedding QE, selection-based explicit QE, and generative QE, highlighting how PLMs/LLMs reshape each paradigm.
    \item[\textemdash] \textit{Practice and impact.} We connect methods to deployment concerns, compare traditional vs. neural QE, and review applications.
\end{itemize}


Section~\ref{sec:foundations_tradQE} revisits the foundations of query expansion and traditional QE methods.
Section~\ref{sec:plm_landscape} overviews the pre-trained model landscape, distinguishing model architectures from training paradigms and their implications for QE.
Section~\ref{sec:plmllm_qe} organizes PLM/LLM-driven QE techniques along four dimensions: Point of Injection, Grounding and Interaction, Learning and Alignment, and KG-Augmented QE.
Section~\ref{sec:applications} surveys application domains and representative use cases.
Section~\ref{sec:blueprint} synthesizes the survey into a deployment-oriented comparative blueprint and a decision framework for selecting and combining QE techniques under real-world constraints.
Section~\ref{sec:future} discusses open challenges and future directions.
Section~\ref{sec:conclusion} concludes the survey.

Figure~\ref{fig:taxonomy} provides an overview of the taxonomy used throughout this survey, spanning traditional QE, PLM/LLM-driven QE techniques, and major application domains.

\newcommand{\TaxoFont}{\fontsize{7.2}{9}\selectfont\normalfont}
\newcommand{\LB}{\(\lbrack\)}
\newcommand{\RB}{\(\rbrack\)}
\definecolor{softblue}{RGB}{214,224,238}
\definecolor{softgreen}{RGB}{218,234,214}
\definecolor{softpurple}{RGB}{220,215,232}
\definecolor{softyellow}{RGB}{246,233,196}
\definecolor{softred}{RGB}{234,210,210}
\definecolor{softgray}{RGB}{228,230,233}
\definecolor{softgold}{RGB}{238,218,182}

\definecolor{softblueD}{RGB}{182,198,220}
\definecolor{softgreenD}{RGB}{188,211,186}
\definecolor{softpurpleD}{RGB}{193,186,209}
\definecolor{softyellowD}{RGB}{232,212,164}
\definecolor{softredD}{RGB}{214,180,180}
\definecolor{softgrayD}{RGB}{206,210,216}
\definecolor{softgoldD}{RGB}{218,192,152}

\definecolor{softblueL}{RGB}{226,234,246}
\definecolor{softgreenL}{RGB}{232,245,230}
\definecolor{softpurpleL}{RGB}{234,230,244}
\definecolor{softyellowL}{RGB}{250,240,212}
\definecolor{softredL}{RGB}{245,228,228}
\definecolor{softgrayL}{RGB}{239,240,242}
\definecolor{softgoldL}{RGB}{245,232,206}

\tikzset{
  taxoShadow/.style={
    blur shadow={
      shadow blur steps=6,
      shadow xshift=0.8pt,
      shadow yshift=-0.8pt,
      shadow opacity=0.16
    }
  },
  leaf/.style={
    draw=black,
    rounded corners, minimum height=1em,
    text width=26em, 
    text opacity=1, align=left,
    fill opacity=.3, text=black, font=\TaxoFont,
    inner xsep=5pt, inner ysep=3pt,
    taxoShadow
  },
leafShort/.style={leaf, text width=17em},
leafLong/.style={leaf, text width=36em},
  leaf1/.style={ 
    draw=black,
    rounded corners, minimum height=1em,
    text width=6.5em, 
    text opacity=1, align=center,
    fill opacity=.5, text=black, font=\TaxoFont,
    inner xsep=3pt, inner ysep=3pt,
    taxoShadow
  },
  leaf2/.style={ 
    draw=black,
    rounded corners, minimum height=1em,
    text width=8em, 
    text opacity=1, align=center,
    fill opacity=.8, text=black, font=\TaxoFont,
    inner xsep=3pt, inner ysep=3pt,
    taxoShadow
  },
  leaf3/.style={ 
    draw=black,
    rounded corners, minimum height=1em,
    text width=7em, 
    text opacity=1, align=center,
    fill opacity=.8, text=black, font=\TaxoFont,
    inner xsep=3pt, inner ysep=3pt,
    taxoShadow
  }
}

\begin{figure*}[ht]
\centering
\scalebox{0.95}{
\begin{forest}
  for tree={
    forked edges,
    grow=east,
    reversed=true,
    anchor=center,
    parent anchor=east,
    child anchor=west,
    base=center,
    font=\TaxoFont,
    rectangle,
    draw=black,
    edge=black!50,
    rounded corners,
    minimum width=2em,
    s sep=3pt,  
    l sep=12pt, 
    inner xsep=3pt,
    inner ysep=2pt
  },
  [{Query Expansion},rotate=90,anchor=north,inner xsep=8pt,inner ysep=5pt,edge=black!50,draw=black, fill=softgrayD
    [{Traditional \\ Query Expansion \\ \S~\ref{sec:foundations_tradQE}}, edge=black!50, leaf3, fill=softgreenD
      [{Global Analysis \\ \S \ref{sec:Global_Analysis}}, leaf2, fill=softgreen
        [{Corpus-Wide Statistical Methods}, leaf1, fill=softgreen
          [{\textbf{\textit{Term Co-occurrence and Clustering}}: ~\cite{minker1972evaluation, willett1988recent, peat1991limitations}}, leafShort, fill=softgreenL]
          [{\textbf{\textit{Global Similarity Thesaurus}}: ~\cite{qiu1993concept}}, leafShort, fill=softgreenL]
        ]
        [{Pre-defined \\ Knowledge Resource}, leaf1, fill=softgreen
          [{\textbf{\textit{General-purpose Lexicons}}: ~\cite{voorhees1993using}}, leafShort, fill=softgreenL]
          [{\textbf{\textit{Domain Ontologies}}: ~\cite{bhogal2007review, lu2009evaluation, aronson1997query, fu2005ontology}}, leafShort, fill=softgreenL]
        ]
      ]
      [{Local Analysis\\ \S \ref{sec:Local_Analysis}}, leaf2, fill=softgreen
        [{\textbf{\textit{PRF/RF}}: Rocchio \cite{rocchio1971relevance}, RSJ weighting \cite{robertson1976relevance}}, leaf, fill=softgreenL]
        [{\textbf{\textit{LCA}}: \cite{xu2000improving, xu2017quary}}, leaf, fill=softgreenL]
      ]
      [{Query Logs \& \\ User Behavior \\ \S \ref{sec:Query_Logs}}, leaf2, fill=softgreen
        [{\textbf{\textit{Mining reformulations or Click-based semantic proximity}}:\\ ~\cite{wen2002query, cui2002probabilistic, cui2003query, baeza2004query, yin2009query}}, leaf, fill=softgreenL]
      ]
    ]
    [{PLM/LLM-Driven \\ QE Techniques\\ \S \ref{sec:plmllm_qe}}, edge=black!50, leaf3, fill=softgoldD
      [{Point of \\ Injection\\ \S \ref{subsec:Point_of_injection}}, leaf2, fill=softgold
        [{\textbf{\textit{Implicit Embedding-based QE}}: ANCE-PRF~\cite{yu2021improving}{,} ColBERT-PRF~\cite{wang2023colbert}{,} Eclipse~\cite{d2025eclipse}{,} QB-PRF~\cite{zhang2024selecting}{,} LLM-VPRF~\cite{li2025llm}}, leaf, fill=softgoldL]
        [{\textbf{\textit{Selection-based Explicit QE}}: CEQE~\cite{naseri2021ceqe}{,} SQET~\cite{naseri2022ceqe}{,} BERT-QE~\cite{zheng2020bert}{,} CQED~\cite{Khader2023LearningTR}{,} PQEWC~\cite{bassani2023personalized}}, leaf, fill=softgoldL]
      ]
      [{Grounding \& \\ Interaction\\ \S \ref{subsec:grounding_interaction}}, leaf2, fill=softgold
        [{\textbf{\textit{Zero-Grounding, Non-Interactive QE: LLM-Driven Single-Stage Expansion}}: Query2Doc~\cite{wang-etal-2023-query2doc}{,} CoT-QE~\cite{jagerman2023query}{,} GAR~\cite{mao2021generation}{,} GRF~\cite{mackie2023generative}{,} HyDE~\cite{gao2023precise}{,} Exp4Fuse~\cite{liu-zhang-2025-exp4fuse}{,} Contextual clue sampling~\cite{liu2022query}{,} SEAL~\cite{bevilacqua2022autoregressive}{,} SU-RankFusion~\cite{li2026dual}{,} Two-LLM QE~\cite{li2026automatic}}, leaf, fill=softgoldL]
        [{\textbf{\textit{Grounding-Only, Non-Interactive QE: Corpus-Evidence Anchored Single-Pass Expansion}}: MILL~\cite{jia2024mill}{,} AGR~\cite{chen2024analyze}{,} EAR~\cite{chuang2023expand}{,} GenPRF~\cite{wang2023generative}{,} CSQE~\cite{lei2024corpus}{,} MUGI~\cite{zhang2024exploring}{,} PromptPRF~\cite{li2025pseudo}{,} FGQE~\cite{jaenich2025fair}}, leaf, fill=softgoldL]
        [{\textbf{\textit{Grounding-Aware Interactive QE: Multi-Round Retrieve-Expand Loops}}: InteR~\cite{feng2024synergistic}{,} ProQE~\cite{rashid2024progressive}{,} LameR~\cite{shen-etal-2024-retrieval}{,} ThinkQE~\cite{lei-etal-2025-thinkqe}}, leaf, fill=softgoldL]
      ]
      [{Learning \& Alignment \\ for QE with PLMs/LLMs\\ \S \ref{subsec:aligned_qe}}, leaf2, fill=softgold
        [{SoftQE~\cite{pimpalkhute2024softqe}{,} RADCoT~\cite{lee2024radcot}{,} ExpandR~\cite{yao2025expandr}{,} AQE~\cite{yang2025aligned}}, leaf, fill=softgoldL]
      ]
      [{KG-Augmented \\ Query Expansion\\ \S \ref{subsec:KG-augmented}}, leaf2, fill=softgold
        [{\textbf{\textit{Entity-based Expansion via Knowledge Graph Retrieval}}: KGQE~\cite{perna2025knowledge}{,} CL-KGQE~\cite{rahman2019query}}, leaf, fill=softgoldL]
        [{\textbf{\textit{Entity-based Expansion via Hybrid KG and Document Graph}}: KAR~\cite{xia-etal-2025-knowledge}{,} QSKG~\cite{mackie2022query}}, leaf, fill=softgoldL]
      ]
    ]
    [{Application \\ Domains\\ \S \ref{sec:applications}}, edge=black!50, leaf3, fill=softpurpleD
      [{\textbf{\textit{Web Search Engines}}: EQE~\cite{zhang2023event}}, leafLong, fill=softpurpleL]
      [{\textbf{\textit{Biomedical Information Retrieval}}: BioBERT/SciBERT~\cite{kelly2021enhancing, khader2022contextual}{,}  BioRAGent\cite{ateia2025bioragent}{,} Re-KGR\cite{niumitigating}}, leafLong, fill=softpurpleL]
      [{\textbf{\textit{E-commerce Search}}: BEQUE~\cite{peng2024large}{,} ~\citet{wang2024leveraging}{,} ~\citet{zhang2022advancing}}, leafLong, fill=softpurpleL]
      [{\textbf{\textit{Cross-Lingual and Multilingual Search}}: \citet{rajaei2024enhancing}}, leafLong, fill=softpurpleL]
      [{\textbf{\textit{Open-Domain Question Answering}}: GAR~\cite{mao2021generation}{,} EAR~\cite{chuang2023expand}{,} AQE~\cite{yang2025aligned}{,} MILL~\cite{jia2024mill}{,} AGR~\cite{chen2024analyze}{,} HTGQE~\cite{zhu2023hybrid}{,} TDPR~\cite{li5367307query}}, leafLong, fill=softpurpleL]
      [{\textbf{\textit{Retrieval-Augmented Generation (RAG)}}: QE-RAGpro~\cite{huquery}{,} QOQA~\cite{koo2024optimizing}{,} KG-Infused RAG~\cite{wu2025kg}{,} ~\citet{rajaei2024enhancing}}, leafLong, fill=softpurpleL]
      [{\textbf{\textit{Conversational Search}}: PRL~\cite{mo2023learning}{,} MISE~\cite{kumar2020making}{,} QuReTeC~\cite{voskarides2020query}}, leafLong, fill=softpurpleL]
      [{\textbf{\textit{Code Search}}: GACR~\cite{li2022generation}{,} SSQR~\cite{mao2023self}{,} SG-BERT/GPT2~\cite{liu2023self}{,} ECSQE~\cite{bibi2025enhancing}}, leafLong, fill=softpurpleL]
    ]
  ]
\end{forest}
}

\caption{A Taxonomy of Query Expansion Techniques: From Traditional Methods to PLM/LLM-Driven Techniques and Applications.}
\label{fig:taxonomy}
\end{figure*}


\section{Foundations of Query Expansion: Concepts and Traditional Methods}
\label{sec:foundations_tradQE}

QE mitigates vocabulary mismatch by adding or reweighting expansion signals so that a query better matches relevant documents \citep{bhogal2007review,sartori2009comparison,efthimiadis1996query,rivas2014study,raza2019survey,lei2011survey,azad2019query}. This section revisits core concepts and traditional QE methods that remain widely used as strong baselines and building blocks in modern pipelines.

\subsection{Concepts and Notation}
Let the original query be a multiset of terms $Q$, stopwords $T''\!\subseteq\!Q$, and a set of candidate expansions $T'$ from some source $D$. Following \citet{azad2019query}, the expanded query is
\begin{equation}
Q_{\text{exp}}=(Q-T'')\ \cup\ T'.
\end{equation}
However, this set-based view is a simplification. In practice, QE is not merely about adding terms but also about re-weighting them to reflect their importance. The core difficulty is selecting $T'$ that improves recall without distorting intent \citep{krovetz1992lexical}. Throughout, we write $D_k$ for top-$k$ feedback documents, and $\cos(\cdot,\cdot)$ is cosine similarity.

\subsection{Global Analysis: Corpus-Wide and External Knowledge}
\label{sec:Global_Analysis}
Global methods operate on the assumption that stable, useful term associations can be inferred from large-scale evidence, independent of any single query. This evidence can be statistical patterns from the entire document corpus or curated knowledge from external resources.
\subsubsection{Corpus-Wide Statistical Methods}

Corpus-wide statistical methods assume that term associations can be inferred from regularities in the target corpus, without requiring explicit relevance judgments.

A classic line of work expands queries using document-level term co-occurrence (e.g., cosine-style association based on how often two terms appear in the same documents). Although simple and unsupervised, such measures have a known bias: they tend to favor terms with document frequencies similar to the query term, which can be weak discriminators under RSJ theory \citep{peat1991limitations,robertson1976relevance}. As a result, co-occurrence expansion is often most beneficial for low-frequency query terms and may be less reliable for high-frequency ones.

Beyond pairwise co-occurrence, global approaches build a corpus-wide thesaurus of term-to-term similarity from term--document statistics \citep{qiu1993concept}. Expansion then selects terms that are globally similar to the query terms (aggregated by query term weights), producing a concept-consistent list. Latent Semantic Analysis (LSA) follows a related idea by mapping terms into a low-dimensional latent space so that semantically related terms become closer \citep{abdelali2007improving}. These global methods avoid hard clustering and support variable-length expansions, but they are largely context-agnostic and may propagate popular yet overly generic terms.

\subsubsection{Methods Using Pre-defined Knowledge Resources}
Curated resources inject explicit semantics and reduce ambiguity \citep{nasir2019knowledge}.
General-purpose Lexicons (e.g., WordNet) offers synsets and IS-A hierarchies for broad-coverage synonymy/hypernymy. However, short queries lack context for reliable WSD; naive synset expansion can underperform stemming/bag-of-words \citep{voorhees1993using,bhogal2007review}. Using glosses or intersecting sense neighborhoods can help \citep{navigli2003analysis}, but static coverage and sense ambiguity remain key issues for complex queries. Domain ontologies (e.g., UMLS/MeSH in biomedicine; legal/tourism ontologies) enable precise, concept-level expansion with fewer sense conflicts \citep{bhogal2007review}. Examples include mapping ``heart attack'' to ``myocardial infarction'' and related treatments/symptoms \citep{lu2009evaluation,aronson1997query}; composing tourism and geographic ontologies to resolve spatial language \citep{fu2005ontology}. Benefits are high precision and expert semantics; costs are curation/maintenance and domain dependence.

\subsection{Local Analysis: Query-Dependent Evidence}
\label{sec:Local_Analysis}
In contrast to global methods, local analysis techniques derive expansion terms from a small set of documents retrieved in response to the current query. This approach is inherently context-aware but is highly dependent on the quality of the initial search results.

\subsubsection{Relevance Feedback (RF) and Pseudo-Relevance Feedback (PRF)}
RF adapts the query using assessed relevant/non-relevant documents; PRF automates this by assuming top-$k$ are relevant.

RF relies on explicitly labeled document sets to refine the query. Given a set of relevant documents $D_r$ and an optional set of non-relevant documents $D_{nr}$, Rocchio \citep{rocchio1971relevance} proposes a vector-space update rule to construct the modified query vector:
\begin{equation}
\vec{q}_m=\alpha \vec{q}_0 + \beta \frac{1}{|D_r|}\!\!\sum_{\vec{d}\in D_r}\!\vec{d}\; -\; \gamma \frac{1}{|D_{nr}|}\!\!\sum_{\vec{d}\in D_{nr}}\!\vec{d},
\end{equation}

where $\alpha, \beta, \gamma$ are non-negative weighting parameters: $\alpha$ controls the influence of the original query vector $\vec{q}_0$, $\beta$ adjusts the contribution of positive evidence from $D_r$, and $\gamma$ modulates the impact of negative evidence from $D_{nr}$. For PRF, only the positive centroid is used, with $D_r$ approximated by the top-$k$ initially retrieved documents $D_k$ as pseudo-relevant feedback.

\paragraph{Probabilistic (RSJ) weighting}
Alternatively, expansion weights follow the RSJ discrimination:
\begin{equation}
w(t)=\log \frac{P(t\mid R)\,[1-P(t\mid NR)]}{P(t\mid NR)\,[1-P(t\mid R)]},
\end{equation}
prioritizing terms overrepresented in relevant documents \citep{robertson1976relevance}. In PRF, $P(t\!\mid\!R)$ is estimated on $D_k$ and $P(t\!\mid\!NR)$ against the background.

\subsubsection{Local Context Analysis (LCA)}

LCA is a query expansion method built on PRF, designed to mitigate topic drift by enforcing local contextual consistency \citep{xu2000improving,xu2017quary}. LCA first takes the top-$n$ initially retrieved documents as a local pseudo-relevant set $S$, and evaluates each candidate concept (a term or phrase) by its degree of co-occurrence with the original query terms within $S$. This estimation combines length-normalized term frequency signals with inverse document frequency (IDF) to emphasize more discriminative concepts. LCA then aggregates the co-occurrence evidence across multiple query terms using a product-based “soft-AND” scheme (with a small smoothing constant to avoid zero effects), thereby favoring candidates supported by most query terms rather than by a single term. Concepts are finally ranked by the resulting score, and the top-$k$ concepts are appended to the original query for expansion. Owing to its emphasis on consistent support across query terms, LCA is generally more robust to topic drift than frequency-only PRF variants.

\subsection{Methods Using Query Logs and User Behavior}
\label{sec:Query_Logs}
At web scale, logs capture implicit reformulations and intent signals \citep{wen2002query,cui2002probabilistic,cui2003query,baeza2004query,yin2009query}.

For a user's original query $Q$, if a subsequent query reformulation $Q'$ frequently follows $Q$ and leads to successful user interactions (e.g., clicks, long dwell time), terms $t$ that appear in $Q'$ but not in $Q$ are treated as expansion candidates \citep{cui2002probabilistic, cui2003query}. To quantify the association between $Q$ and each candidate $t$, we use a statistic assoc(Q,t) that calculates the proportion of $Q$'s submissions where the subsequent $Q'$ contains $t$:
\begin{align}
\mathrm{assoc}(Q, t)=\frac{\mathrm{count}\big(Q  \!\to\! Q' \text{ with } t \!\in\! Q'\big)}{\mathrm{count}(Q)}.
\end{align}

Queries leading to overlapping clicked document sets are semantically related. Let $D_Q$ denote the historical clicked set of a query $Q$ (i.e., all documents clicked by users after submitting $Q$); the semantic similarity between two queries $Q_1$ (to be expanded) and $Q_2$ is calculated as:
\begin{align}
\mathrm{sim}(Q_1,Q_2)=\frac{|D_{Q_1}\cap D_{Q_2}|}{|D_{Q_1}\cup D_{Q_2}|},
\end{align}
and terms from $Q_2$ (weighted by $\mathrm{sim}$ and within-$Q_2$ importance) can expand $Q_1$. These approaches are strongly data-driven but suffer from cold-start/long-tail sparsity and privacy constraints.

\subsection{Limitations of Traditional QE}
\label{sec:limitations_of_Traditional_QE}
Although diverse in their approach, traditional QE techniques are bound by a shared, fundamental limitation: they operate on surface-level lexical statistics or static knowledge structures, lacking a deep model of dynamic, compositional semantics. This core deficiency manifests in several ways:

\vspace{0.5em} 
\begin{itemize}
    \setlength{\leftmargin}{1.5em}
    \setlength{\labelwidth}{0.5em}
    \setlength{\labelsep}{0.5em}
    \item[\textemdash] \textit{Context Insensitivity}
    Global methods, whether statistical or knowledge-based, are inherently context-agnostic. WordNet's failure to resolve ambiguity in short queries is a direct consequence of this, as the correct "sense" of a word is determined by its surrounding terms—a signal these models cannot effectively use.

    \item[\textemdash] \textit{Brittleness of Local Evidence}
    Local methods like PRF are highly susceptible to topic drift because they treat documents as bags of co-occurring words. They lack the ability to understand the compositional intent of a query like "running shoes" and may drift towards the more general and popular topic of "shoes" if the initial results are noisy.

    \item[\textemdash] \textit{Inability to Generalize}
    Log-based methods are powerful but cannot generalize to novel queries. They rely on observing past user behavior for specific query patterns and fail when encountering new combinations of terms that express a previously unseen intent.
\end{itemize}

In sum, traditional QE is constrained by context-agnostic associations and limited generalization beyond observed evidence, which hampers disambiguation, robustness, and long-tail coverage. These limitations motivate PLM/LLM-era approaches that learn contextual representations and support controllable, evidence-aware expansion (see Sec.~\ref{sec:plmllm_qe}).

\section{Pre-trained Model Landscape}
\label{sec:plm_landscape}

Modern QE increasingly leverages PLMs/LLMs. This section summarizes the model families most relevant to QE and the capabilities they expose (contextual encoding, controlled generation, and instruction following), while deferring concrete QE techniques to Sec.~\ref{sec:plmllm_qe}.

\subsection{Model Architectures}
\label{subsec:model_architectures}
\subsubsection{Encoder-only}
\label{subsubsec:enc_only}
Encoder-only transformers (e.g., BERT, RoBERTa) are bidirectional masked LMs trained to produce context-sensitive token/sequence representations \citep{devlin2019bert,liu2019roberta}. For QE, their strengths are:

\vspace{0.5em} 
\begin{itemize}
    \setlength{\leftmargin}{1.5em}
    \setlength{\labelwidth}{0.5em}
    \setlength{\labelsep}{0.5em}
    \item[\textemdash] \textit{Context-sensitive disambiguation.}
    For queries like ``Apple launch event'', encoder-only models resolve sense and yield embeddings that align with Apple Inc. rather than the fruit, enabling safer selection-based expansion (e.g., synonym/phrase picking) instead of blind synonym injection.

    \item[\textemdash] \textit{Implicit/embedding-space expansion.}
    They naturally support representation augmentation (e.g., PRF over dense vectors, token-level reweighting) without generating surface forms, which mitigates hallucination and reduces latency.
\end{itemize}

These models are a strong fit for PRF-style, selection-based, or reranking-centric QE where precision and efficiency are paramount (see Sec.~\ref{sec:plmllm_qe}).
\subsubsection{Encoder--Decoder}
\label{subsubsec:enc_dec}
Sequence-to-sequence (seq2seq) models pair a bidirectional encoder with an autoregressive decoder, combining understanding and generation. BART implements denoising autoencoding with span permutation/masking, synthesizing BERT-like encoding and GPT-like decoding \citep{Lewis2019BARTDS}. T5 unifies tasks under a text-to-text objective and scales with large pre-training corpora \citep{Raffel2019ExploringTL}. For QE, encoder--decoders offer:

\vspace{0.5em} 
\begin{itemize}
    \setlength{\leftmargin}{1.5em}
    \setlength{\labelwidth}{0.5em}
    \setlength{\labelsep}{0.5em}
    \item[\textemdash] \textit{Direct generation of expansions.}
    They produce answer-like pseudo-documents, titles, or rationales that enrich sparse/dense retrieval.

    \item[\textemdash] \textit{Structure-aware control.}
    Prompts or templates can steer outputs to keywords, entities, or facets, improving interpretability and reducing topic drift.
\end{itemize}

Trade-offs include decoding latency and the need to control generation quality; however, they remain the dominant backbone for generative QE methods.

\subsubsection{Decoder-only LLMs}
\label{subsubsec:dec_only}
Decoder-only transformers (e.g., GPT-3/4, PaLM, LLaMA-family) are trained autoregressively to predict the next token on massive corpora \citep{achiam2023gpt,touvron2023llama,team2023gemini}. This architecture serves as a powerful generative backbone, scaling effectively to exhibit \textit{emergent} capabilities like in-context learning \citep{weiemergent}. For QE, the architectural affordances are:

\vspace{0.5em} 
\begin{itemize}
    \setlength{\leftmargin}{1.5em}
    \setlength{\labelwidth}{0.5em}
    \setlength{\labelsep}{0.5em}
    \item[\textemdash] \textit{Generative Capacity.}
    Unlike encoder-only models that extract or re-weight existing terms, the decoder-only architecture can synthesize entirely new text sequences, such as pseudo-documents or hypothetical answers, providing rich context for retrieval \citep{wang-etal-2023-query2doc}.

    \item[\textemdash] \textit{In-context Adaptation (Few-shot).}
    Due to their autoregressive nature, these models can perform query expansion by mimicking demonstrations provided in the prompt (few-shot prompting), effectively learning the expansion pattern without weight updates.
\end{itemize}

It is important to note that while the decoder-only architecture provides the necessary capacity, the ability to follow complex zero-shot expansion instructions (e.g., reasoning-guided reformulation) is primarily unlocked by the \textit{instruction tuning} paradigm, which we discuss in Sec.~\ref{subsec:training_paradigms}.


\subsection{Training Paradigms: Unlocking Capabilities}
\label{subsec:training_paradigms}
While model architectures determine the fundamental inductive biases (e.g., generation vs. encoding), it is the training paradigm that elicits specific capabilities required for effective QE. We categorize the training pipeline into three progressive stages: pre-training, instruction tuning, and alignment.
\subsubsection{Pre-training: Knowledge Acquisition}
\label{subsubsec:pretraining}
Pre-training on massive corpora provides the model with world knowledge and linguistic distributions via objectives like masked language modeling (MLM) or causal language modeling (CLM) \citep{devlin2019bert,brown2020language}. 
For QE, this stage provides semantic association. Base models capture co-occurrence patterns (e.g., associating ``renal failure'' with ``kidney disease''), which underpins embedding-based expansion. However, base CLM models (e.g., GPT-3) function primarily as text completers and often struggle to follow explicit expansion instructions without further tuning.
\subsubsection{Instruction Tuning: Eliciting Reasoning}
\label{subsubsec:sft}
Supervised Fine-Tuning (SFT) on diverse instruction--response pairs is the critical step that unlocks zero-shot generalization and reasoning capabilities \citep{weifinetuned,ouyang2022training}. Models like FLAN-T5 (encoder-decoder) and LLaMA-Chat (decoder-only) are fine-tuned to map user instructions to desired outputs.
For QE, SFT is transformative:
\vspace{0.5em} 
\begin{itemize}
    \setlength{\leftmargin}{1.5em}
    \setlength{\labelwidth}{0.5em}
    \setlength{\labelsep}{0.5em}

    \item[\textemdash] \textit{Instruction Following.}
It improves adherence to task-specific QE instructions (e.g., ``extract keywords'' or ``generate a hypothetical document''), reducing the tendency to merely continue the input query text.

\item[\textemdash] \textit{Zero-shot Generalization and Structured Reasoning.}
SFT can improve zero-shot transfer to unseen QE tasks and, when trained on reasoning-oriented instruction data, better support structured decomposition of complex queries or explicit justifications for expansion terms.
\end{itemize}

\subsubsection{Alignment (RLHF/DPO): Safety and Control}
\label{subsubsec:alignment}
Following SFT, models often undergo alignment via Reinforcement Learning from Human Feedback (RLHF) or Direct Preference Optimization (DPO) to steer generation toward desired behaviors and constraints \citep{ouyang2022training,rafailov2023direct}.
For QE, this stage primarily improves safety, reliability, and controllability by shaping the model's outputs with task-relevant preference signals:
\vspace{0.5em} 
\begin{itemize}
    \setlength{\leftmargin}{1.5em}
    \setlength{\labelwidth}{0.5em}
    \setlength{\labelsep}{0.5em}

    \item[\textemdash] \textit{Hallucination Control.}
    Preference optimization can discourage the generation of fabricated entities (e.g., fake citations or non-existent medical drugs), especially when the preference data explicitly penalizes unsupported claims. However, alignment alone does not guarantee factual correctness unless factuality is directly rewarded or constrained.

    \item[\textemdash] \textit{Task Helpfulness for Retrieval Intent.}
    By optimizing toward QE-specific preferences (e.g., intent coverage, diversity, and non-redundancy under a keyword-like format), alignment biases the model toward expansions that are practically useful for downstream retrieval, reducing chatty, generic, or repetitive content that may not translate into search gains.

    \item[\textemdash] \textit{Output Controllability.}
    Alignment can improve compliance with generation constraints (e.g., length limits, keyword-only output, avoidance of sensitive or off-intent terms), making QE outputs more predictable and easier to integrate into retrieval pipelines.
\end{itemize}

\subsection{Domain-specific and Multilingual Models}
\label{subsec:domain_multi}
Specialized pre-training narrows vocabulary mismatch in professional and non-English settings:
\vspace{0.5em} 
\begin{itemize}
    \setlength{\leftmargin}{1.5em}
    \setlength{\labelwidth}{0.5em}
    \setlength{\labelsep}{0.5em}
    \item[\textemdash] \textit{Domain-specific encoders.}
    BioBERT adapts BERT on PubMed/PMC for biomedical text \citep{Lee2019BioBERTAP}; UMLS-BERT injects UMLS CUIs/semantic types \citep{Michalopoulos2020UmlsBERTCD}; SciBERT trains on scientific full text with SCIVOCAB to better cover scholarly terminology \citep{Beltagy2019SciBERTAP}. These models consistently improve entity resolution and concept linking, which directly benefits QE in biomedical and scientific IR.

    \item[\textemdash] \textit{Multilingual PLMs/LLMs.}
    Multilingual variants (e.g., multilingual BERT families ~\citep{pan2021multilingual, devlin2019bert}, mT5~\citep{xue2021mt5}) support cross-lingual QE by mapping semantically aligned terms across languages, enabling bilingual/multilingual reformulations without parallel supervision. In practice, they are used to (i) translate-and-expand or (ii) expand-then-translate within cross-lingual retrieval pipelines.
\end{itemize}

\subsection{Architectural Affordances for QE: A Practitioner’s View}
\label{subsec:affordances}

Table~\ref{tab:plm_qe_affordances} summarizes typical QE affordances by model architecture. In addition, the survey highlights two recurring factors that substantially affect deployment behavior across architectures: (i) instruction/pref\-erence tuning, which typically improves output format adherence and controllability, and (ii) domain or multilingual continued pre-training, which helps bridge vocabulary gaps in specialized settings. Sec.~\ref{sec:plmllm_qe} details concrete algorithms that instantiate these affordances.

{\renewcommand{\arraystretch}{1.15}
\begin{table}[h]
\small
\centering
\caption{Pre-trained model architectures and typical QE affordances.}
\label{tab:plm_qe_affordances}
\begin{tabularx}{\textwidth}{p{0.22\textwidth} p{0.22\textwidth} p{0.28\textwidth} p{0.22\textwidth}}
\toprule
\textbf{Architecture} & \textbf{Pre-training gist} & \textbf{QE strengths (typical use)} & \textbf{Common trade-offs} \\
\midrule
Encoder-only (BERT/RoBERTa) \citep{devlin2019bert,liu2019roberta} &
Bidirectional masked LM &
Selection/embedding-space QE (term/span selection, dense PRF); efficient reranking features &
Limited generation; relies on candidate pools / first-stage evidence \\
\addlinespace
Encoder--Decoder (BART/T5) \citep{Lewis2019BARTDS,Raffel2019ExploringTL} &
Seq2seq denoising / text-to-text &
natural for conditional generation / templated outputs (e.g., rewrites or short expansion texts) &
Generation latency; often needs task-specific tuning/data for stable QE quality \\
\addlinespace
Decoder-only LLM (GPT-style/PaLM/LLaMA) \citep{floridi2020gpt,achiam2023gpt,anil2023palm,touvron2023llama} &
Autoregressive LM; in-context learning \citep{weiemergent} &
Prompt-based zero-/few-shot expansion; reasoning-guided reformulation &
Higher generation cost (often longer/multi-candidate outputs); drift/hallucination risk without grounding \\
\bottomrule
\end{tabularx}
\end{table}
}

Instruction/preference tuning \citep{ouyang2022training,weifinetuned,touvron2023llama} and domain/multilingual continued pre-training \citep{Lee2019BioBERTAP,Michalopoulos2020UmlsBERTCD,Beltagy2019SciBERTAP,devlin2019bert,pan2021multilingual} affect controllability and vocabulary coverage across all architectures; recent work has placed particular emphasis on such adaptations for decoder-only LLMs \citep{yang2025qwen3}.

\paragraph{Model access in practice.}
Beyond model architecture, many QE pipelines differ in how the model is accessed and customized.
Hosted LLM services accessed via APIs (e.g., ChatGPT- or Gemini-style \cite{team2023gemini} systems) make prompt-only QE easy to prototype and deploy, but raise practical concerns such as latency/cost budgets, limited transparency, and version drift.
Self-hosted/open-weight models enable tighter integration and adaptation (e.g., PEFT or distillation) at the cost of additional engineering and serving complexity.
These considerations connect naturally to the deployment discussion in Sec.~\ref{sec:blueprint}.

\paragraph{From Model Selection to QE Implementation}
The preceding guidance maps PLM/LLM families to retrieval priorities. Realizing their QE value requires addressing four key questions: how to inject expansion signals, ground expansions in corpus evidence, align generation with retrieval goals, and integrate structured knowledge. Sec.~\ref{sec:plmllm_qe} operationalizes these as four axes to bridge model capabilities with practical techniques, forming a unified framework for neural query expansion. These questions and axes structure the technical discussion in Sec.~\ref{sec:plmllm_qe}.

\section{PLM/LLM-Driven QE Techniques}
\label{sec:plmllm_qe}
Modern query expansion revisits a classic idea---add missing evidence to a query---under the lens of PLMs and LLMs.
What changes in the LLM era is not the goal but where and how the additional signal is injected: into the query’s vector representation, into a curated set of explicit terms or spans, or into freshly produced text that stands in for what the user might have asked if they had been more specific.
Equally important are choices about grounding, i.e., whether expansion is anchored in the corpus or relies on a model’s world knowledge, as well as the learning regime (prompt-only versus alignment and distillation) and the interaction pattern (single shot versus feedback in the loop).

We structure the literature along four complementary axes that unify "traditional" QE with PLM/LLM-based methods.

\vspace{0.5em} 
\begin{itemize}
    \setlength{\leftmargin}{1.5em}
    \setlength{\labelwidth}{0.5em}
    \setlength{\labelsep}{0.5em}
    
    \item[\textemdash] \textit{Point of injection.}
    \begin{itemize}
        \setlength{\leftmargin}{1.2em} 
        \setlength{\labelwidth}{0em}
        \setlength{\labelsep}{0.3em}
        \item \textit{Implicit, embedding-level} methods strengthen the query vector or token embeddings without emitting new terms (e.g., ANCE-PRF, ColBERT-PRF, Eclipse, QB-PRF, LLM-VPRF).
        \item \textit{Selection-based explicit} methods keep the vocabulary corpus-grounded by ranking terms or spans from pseudo-relevant documents or curated resources (e.g., CEQE, SQET, BERT-QE, CQED, PQEWC).
    \end{itemize}

    \item[\textemdash] \textit{Grounding and interaction.}
    \begin{itemize}
        \setlength{\leftmargin}{1.2em}
        \setlength{\labelwidth}{0em}
        \setlength{\labelsep}{0.3em}
        \item \textit{Zero-grounding, non-interactive} approaches rely mainly on an LLM’s prior knowledge, sometimes with minimal corpus hints (e.g., Query2Doc, CoT-QE, HyDE, Exp4Fuse), and issue the expanded query once without additional retrieval--generation cycles.
        \item \textit{Grounding-only, non-interactive} methods explicitly condition generation on pseudo-relevant documents or other corpus evidence in a single retrieve--generate--requery pass, often using constraints, selection, or feedback calibration to control drift (e.g., MILL, AGR, EAR, GenPRF, CSQE, MUGI, PromptPRF, FGQE).
        \item \textit{Grounding-aware interactive} designs run multiple retrieve--expand loops (e.g., InteR, ProQE, ThinkQE), using feedback from earlier retrieval stages to guide later expansions.
    \end{itemize}

    \item[\textemdash] \textit{Learning and alignment.}
    \begin{itemize}
        \setlength{\leftmargin}{1.2em}
        \setlength{\labelwidth}{0em}
        \setlength{\labelsep}{0.3em}
        \item Beyond zero/few-shot prompting, systems \textit{align} generation to retrieval utility via supervised fine-tuning, preference optimization, or distillation (e.g., SoftQE, RADCoT, ExpandR, AQE).
    \end{itemize}

    \item[\textemdash] \textit{Knowledge Graph augmentation.}
    \begin{itemize}
        \setlength{\leftmargin}{1.2em}
        \setlength{\labelwidth}{0em}
        \setlength{\labelsep}{0.3em}
        \item \textit{KG-augmented} strategies are categorized into two types: those using only knowledge graph (KG) signals (e.g., KGQE and CL-KGQE) and those integrating KG with document-level graphs (e.g., KAR and QSKG).
    \end{itemize}
\end{itemize}

To make the relationships among these axes more explicit, Table~\ref{tab:cross_axis_matrix} summarizes representative PLM/LLM-driven QE families across point of injection, grounding and interaction, learning/alignment, and structured knowledge use. Rather than introducing a new taxonomy, the table serves as a reader-oriented map that shows how the axes jointly characterize each method family.

Because the following method descriptions involve both textual and vector-space expansion signals, Table~\ref{tab:notation_sec4} provides a quick reference to the main notation used in Section~4.



\begin{table*}[t]
\centering
\small
\caption{Cross-axis summary of representative PLM/LLM-driven QE categories.}
\label{tab:cross_axis_matrix}
\setlength{\tabcolsep}{6pt}
\renewcommand{\arraystretch}{1.2}
\begin{tabularx}{\textwidth}{
  >{\raggedright\arraybackslash}p{0.23\textwidth}
  >{\raggedright\arraybackslash}p{0.46\textwidth}
  >{\raggedright\arraybackslash}X                  
}
\toprule
\textbf{QE category} &
\textbf{Cross-axis profile} &
\textbf{Takeaway} \\
\midrule

Implicit embedding-based QE &
Latent signal; corpus feedback; mostly non-interactive &
Low drift and latency, but first-pass dependent. \\ \addlinespace

Selection-based explicit QE &
Terms, spans, or chunks; corpus-grounded selection &
Controllable, but candidate-pool limited. \\ \addlinespace

Zero-grounding generative QE &
Generated text; model-prior-based; single-shot retrieval &
Useful for sparse evidence, but risks hallucination or drift. \\ \addlinespace

Grounded / interactive generative QE &
Generated text; retrieved evidence; single- or multi-round loops &
Better corpus alignment, but higher cost. \\ \addlinespace

Learning / alignment QE &
Textual, soft, or latent signals; supervision, preferences, or distillation &
Better retrieval utility, but needs extra training signals. \\ \addlinespace

KG-augmented QE &
Entities, relations, or facts; external or hybrid graph grounding &
Improves disambiguation, but depends on KG coverage. \\

\bottomrule
\end{tabularx}
\end{table*}

\begin{table*}[t]
\centering
\small
\caption{Notation used in PLM/LLM-driven query expansion methods.}
\label{tab:notation_sec4}
\setlength{\tabcolsep}{4pt}
\renewcommand{\arraystretch}{1.08}
\begin{tabularx}{\textwidth}{p{0.12\textwidth} X p{0.12\textwidth} X}
\toprule
\textbf{Notation} & \textbf{Meaning} & \textbf{Notation} & \textbf{Meaning} \\
\midrule
$q$ & Original input query in text form. &
$Q$ & Original query as a set or multiset of terms. \\

$\mathbf{q}$ & Vector representation of the query. &
$q^{+}$ & Expanded or augmented query. \\

$g$ & Generated expansion, e.g., pseudo-document, rationale, answer, or clue. &
$G(q)$ & A set of generated expansions for query $q$. \\

$[q;g]$ & Textual concatenation of the query and generated expansion. &
$T'$ & Candidate expansion terms. \\

$R_k(q)$ / $D_k$ & Top-$k$ feedback documents retrieved for query $q$. &
$D_r$, $D_{nr}$ & Relevant and non-relevant document sets in relevance feedback. \\

$C$ & Candidate pool of terms, spans, chunks, entities, or query variants. &
$c_i$ & A selected chunk or candidate expansion unit. \\

$\mathbf{d}$ & Vector representation of a document. &
$s(q,d)$ & Retrieval or relevance score between query $q$ and document $d$. \\

$\mathrm{sim}(\cdot,\cdot)$ / $\cos(\cdot,\cdot)$ & Similarity function, often cosine similarity. &
$\alpha,\beta,\gamma,\lambda$ & Interpolation, feedback, or fusion weights. \\

$L$ & Language model or large language model used for generation or scoring. &
$K$ / $k$ & Number of feedback items, candidates, centroids, or generated outputs. \\
\bottomrule
\end{tabularx}
\end{table*}

\subsection{Point of injection}
\label{subsec:Point_of_injection}
\begin{figure}[htbp]
    \centering
    \includegraphics[width=0.9\textwidth,trim=10 10 10 10,clip]{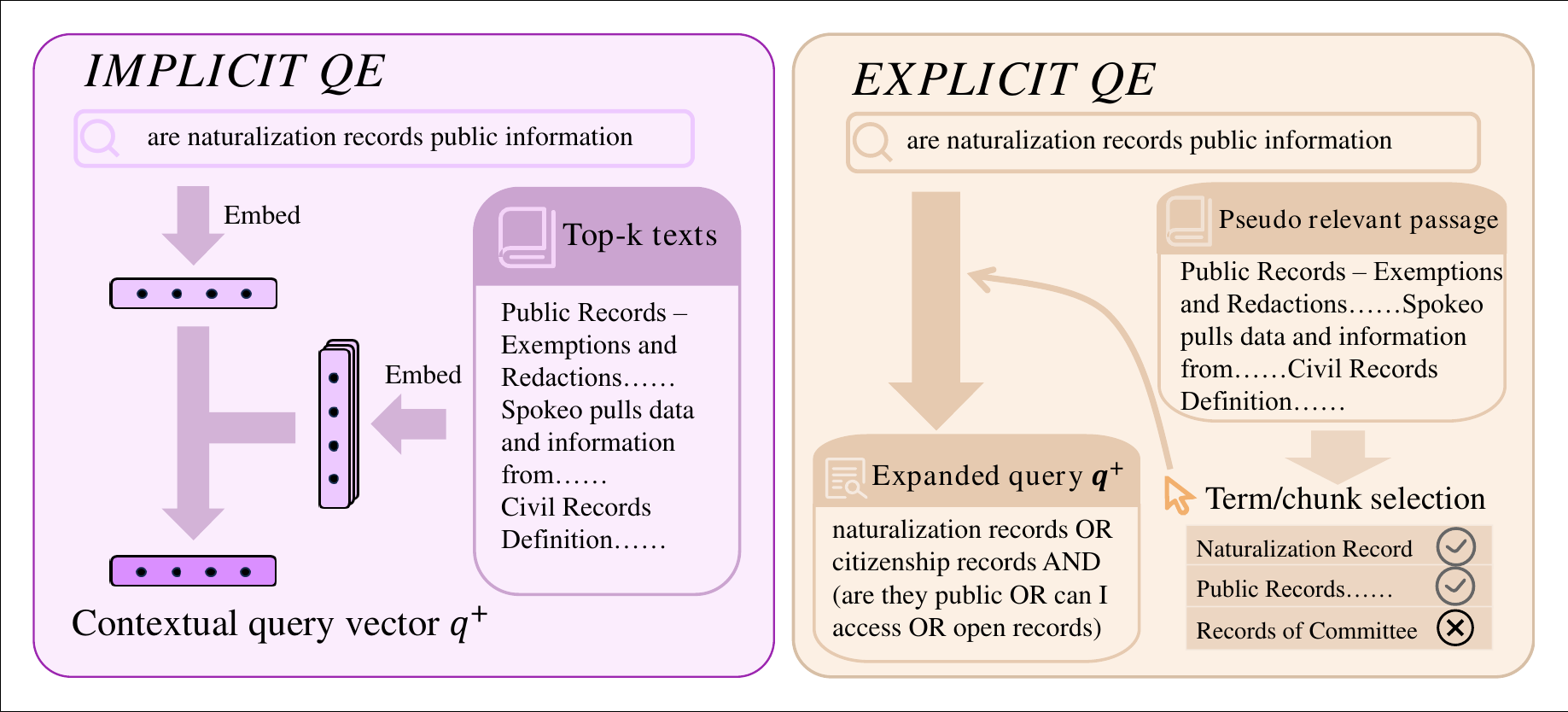}
    \caption{Point of injection in the retrieval pipeline: implicit (embedding-level) vs. explicit (text-level) query expansion.}
    \label{fig:Implicit QE vs Explicit QE}
\end{figure}

Modern QE methods can be roughly categorized into two types, depending on how the expansion signal enters the retrieval pipeline.
On one hand, the query can be implicitly enhanced in the vector space, adjusting its representation to improve retrieval performance. On the other hand, QE can be explicitly performed at the text level, by adding extended terms, spans, or generated text to directly enrich the query content. Clarifying this distinction provides a clear foundation for understanding how and where expansion information can be injected into the retrieval pipeline, which is the focus of this section. Fig.~\ref{fig:Implicit QE vs Explicit QE} illustrates the difference in injection points between Implicit QE and Explicit QE.



\subsubsection{Implicit Embedding-based QE}
\label{subsec:implicit_emb_qe}

Implicit (embedding-level) QE does not emit new terms. Instead, it refines the query representation using signals distilled from a first-pass result set and then re-queries the same index.

Let $q$ be the input query and $\mathcal{R}_k=\{d_1,\dots,d_k\}$ the top-$k$ texts returned by a first pass (when available).
A document $d$ has a fixed embedding $\mathbf{d}\in\mathbb{R}^m$; a query is represented either by a single vector $\mathbf{q}\in\mathbb{R}^m$ (bi-encoder) or by a sequence of token embeddings $\{\boldsymbol{\phi}_{q_i}\}_{i=1}^{|q|}$ (late interaction).

Compared to generative expansion, implicit QE avoids vocabulary drift and decoding latency. It is especially effective when the first pass already surfaces some on-topic material, because the model can amplify directions in embedding space that correlate with relevance and suppress distracting ones.
Below we describe representative methods that implement this idea under single-vector and late-interaction retrievers.

\paragraph{ANCE-PRF: refining a single query vector with PRF}
\textbf{ANCE-PRF}~\cite{yu2021improving} augments a bi-encoder retriever without modifying the document index. 
Given an initial result set $\mathcal{R}_k$, it learns a refined query representation $\mathbf{q}_{\mathrm{PRF}}$ by integrating feedback documents into the original query $q$. 

Retrieval is then performed using the standard inner product between $\mathbf{q}_{\mathrm{PRF}}$ and the fixed document embeddings $\mathbf{d}$ from the original index. 
Training follows a contrastive objective that increases similarity to relevant documents while decreasing similarity to sampled negatives.At inference time, this approach requires one additional encoding pass and a second retrieval step. Its effectiveness depends on $\mathcal{R}_k$ containing at least some relevant content.

\paragraph{ColBERT-PRF: injecting discriminative token directions}
\label{subsubsec:colbertprf}

\textbf{ColBERT-PRF}~\cite{wang2023colbert} adapts pseudo-relevance feedback to late interaction~\cite{Khattab2020ColBERT}. ColBERT scores by matching each query token to its best document token,
$s(q,d)=\sum_{i}\max_{j}\langle \boldsymbol{\phi}_{q_i}, \boldsymbol{\phi}_{d_j}\rangle$.
ColBERT-PRF enriches the query side with a small set of feedback centroids distilled from $\mathcal{R}_k$. It collects token embeddings from $\mathcal{R}_k$ and clusters them into $K$ centroids $\{\boldsymbol{\nu}_\ell\}_{\ell=1}^{K}$, then assigns each centroid a discriminativeness weight
$\sigma_\ell=\log\frac{N+1}{N_\ell+1}$,
where $N_\ell$ is the number of passages containing the nearest lexical token to $\boldsymbol{\nu}_\ell$ (an IDF proxy) and $N$ is the total number of passages in the collection.
The final score appends these centroids as extra ``query tokens'' with weight $\beta$:
\begin{align}
s_{\mathrm{PRF}}(q,d)
=\sum_{i}\max_{j}\langle \boldsymbol{\phi}_{q_i}, \boldsymbol{\phi}_{d_j}\rangle
\;+\;
\beta \sum_{(\boldsymbol{\nu}_\ell,\sigma_\ell)\in F_e} \sigma_\ell\,\max_{j}\langle \boldsymbol{\nu}_\ell,\boldsymbol{\phi}_{d_j}\rangle,
\end{align}
where $F_e$ contains the top-weighted centroids.
These feedback centroids capture discriminative semantic directions observed in $\mathcal{R}_k$ without committing to specific surface forms, which can reduce polysemy-driven drift. In deployment, clustering introduces a modest overhead that can be mitigated with approximate methods, and the approach is most reliable when $\mathcal{R}_k$ is at least partially on-topic; using a small $K$ together with IDF-like weighting helps limit noise imported from off-topic feedback.

\paragraph{Eclipse: reweighting embedding dimensions with positive and negative signals}
\label{subsubsec:eclipse}

\textbf{Eclipse}~\cite{d2025eclipse} stays in the single-vector regime and adjusts dimensions rather than tokens. From the first pass it takes a positive set $D^{+}$ (top results) and a negative set $D^{-}$ (tail results). For each embedding dimension $r\in\{1,\dots,m\}$ it estimates an importance
\begin{align}
\mathrm{imp}(r)
=\frac{\mathrm{avg}_{d\in D^{+}}\,v_{d,r}-\mathrm{avg}_{d\in D^{-}}\,v_{d,r}}{\sigma_r+\epsilon},
\end{align}
where $v_{d,r}$ is the $r$-th coordinate of document vector $\mathbf{v}_d$, $\sigma_r$ is the standard deviation of that coordinate across $D^{+}\cup D^{-}$, and $\epsilon$ stabilises the ratio.
The augmented query is a Hadamard (elementwise) reweighting,
$\mathbf{q}_{\mathrm{aug}}=\mathbf{q}\odot \mathrm{imp}$,
and the second pass uses $\langle \mathbf{q}_{\mathrm{aug}},\mathbf{d}\rangle$.
This reweighting amplifies dimensions that consistently separate positives from negatives and dampens dimensions activated by off-topic material, acting as a soft, data-driven denoiser in latent space. It adds no extra encoding and only requires simple statistics over $D^{+}$ and $D^{-}$; in practice, choosing $D^{-}$ too aggressively can over-suppress useful but rare signals, so small negative pools together with the standard-deviation normalization help mitigate this risk.

\paragraph{QB-PRF: Query-Bag Pseudo-Relevance Feedback (embedding-level)}
\textbf{QB-PRF}~\cite{zhang2024selecting} augments the query representation using a small set of semantically equivalent query variants selected from an initial retrieval stage, and then fuses them in embedding space. Let $q$ be the input query and $f(\cdot)$ the query encoder used by the retriever. A first pass returns a candidate pool $\mathcal{C}(q)=\{q'_1,\dots,q'_M\}$ (e.g., related user queries or automatically mined paraphrases). 

Candidates are embedded and filtered for semantic equivalence and lexical diversity to form a query bag $\mathcal{B}(q)\subset\mathcal{C}(q)$. In practice, a contrastive encoder or variational autoencoder scores equivalence; we denote the retained items as $\{q'_i\}_{i=1}^n$.

Let $\mathbf{q}=f(q)$ and $\mathbf{q}'_i=f(q'_i)$. QB-PRF produces an augmented representation by attentional pooling over the bag: each variant $q'_i$ receives a weight $a_i$ via a softmax over its similarity to $\mathbf{q}$ (with temperature $\tau$), and the final vector interpolates the original query with the weighted average of the bag, controlled by $\beta\in[0,1]$. The second pass scores documents with $s(d\mid q^{+})= \mathbf{q}_{\text{aug}}^{\top}\mathbf{d}$ (for dual-encoders) or by replacing the query side in a late-interaction scorer (for multi-representation models).

QB-PRF operates entirely in the latent space: it does not append tokens nor modify the index, and it complements classical PRF by injecting paraphrastic evidence that is robust to lexical mismatch. Reported results show consistent gains in both early precision and deep recall for conversational and ad hoc retrieval settings, with negligible online cost beyond one additional query encoding.

\paragraph{LLM-VPRF: Vector Pseudo-Relevance Feedback for LLM-based retrievers}
\textbf{LLM-VPRF}~\cite{li2025llm} extends pseudo-relevance feedback to dense retrievers whose query encoders are derived from large language models, such as PromptReps, RepLLaMA, and LLM2Vec. Starting from an initial query vector $\mathbf{q}_0=f(q)$, a first pass retrieves the top-$k$ documents $\mathcal{R}_k(q)=\{d_1,\dots,d_k\}$ with embeddings $\{\mathbf{d}_j\}$ and scores $\{s_j\}$. LLM-VPRF updates the query by interpolating the original vector with a weighted average of the feedback document embeddings:
\begin{align}
\mathbf{q}_{\mathrm{vprf}} = (1-\alpha)\,\mathbf{q}_0 + \alpha\,\frac{\sum_{j=1}^k w_j\,\mathbf{d}_j}{\sum_{j=1}^k w_j},
\end{align}
where $\alpha\in[0,1]$ controls the contribution of feedback, and $w_j$ can be uniform, score-based (e.g., a softmax over $\{s_j\}$), or IDF-based.
The refined query vector is then used for a second retrieval pass on the same index, without text generation, tokenization changes, or model fine-tuning. This approach incorporates corpus-specific semantics from the initial results while remaining lightweight. Experiments on MS~MARCO~\cite{DBLP:conf/nips/NguyenRSGTMD16}, BEIR~\cite{thakur2021beir}, and LoTTE~\cite{santhanam2022colbertv2} report consistent gains in nDCG@10 and Recall@1000 over zero-shot LLM retrievers, with larger improvements on queries with low first-pass recall, and it can be integrated into dual-encoder pipelines alongside fusion or re-ranking methods.



\subsubsection{Selection-based Explicit QE}
\label{subsec:selection_explicit}

Let $q$ denote the input query and $\mathcal{R}_k=\{D_1,\dots,D_k\}$ the pseudo-relevant set returned by a first pass (e.g., BM25 or a dense retriever).
From $\mathcal{R}_k$ we derive a pool of candidate units $C$ to consider for expansion. A candidate can be a term $t$, an $n$-gram, or a short chunk $c$.
We write $\text{sim}(\mathbf{a},\mathbf{b})$ for cosine similarity between vectors $\mathbf{a}$ and $\mathbf{b}$, and $\langle\mathbf{a},\mathbf{b}\rangle$ for an inner product.
Selection-based explicit QE uses a pretrained language model (PLM) not to generate text, but to score candidates in context and keep only those that are likely to help retrieval; the chosen items are then inserted into the query or used as additional evidence at scoring time.
This design gives tight control over vocabulary and makes it easy to incorporate domain or user constraints.

The core idea of selection-based explicit QE with PLMs is to identify salient terms or chunks from the input query and incorporate them into the reformulated query. Building on this principle, several representative approaches have been proposed, which differ in how they select informative units and integrate them with the original query. In the following, we present some of these representative methods in detail.

\paragraph{CEQE: Context-aware scoring in an RM3 pipeline}

\textbf{CEQE}~\cite{naseri2021ceqe} replaces the frequency counts in RM3 with context-sensitive similarities computed by BERT. 
Given an initial feedback set $\mathcal{R}_k$, the query is encoded into a vector $\mathbf{q}$. 
For each document $D \in \mathcal{R}_k$ and each term $t$ occurring in $D$, CEQE extracts contextual embeddings and measures their similarity to the query.

A term $t$ is weighted according to how strongly its contextual occurrences align with $\mathbf{q}$, normalized within each document. 
These document-level signals are aggregated across all feedback documents using standard RM3 weighting $p(q \mid D)$, yielding a feedback model 
\begin{align}
p(t \mid \theta_{\mathcal{R}}) \propto \sum_{D \in \mathcal{R}_k} \text{score}(t, D)\, p(q \mid D),
\end{align}
where $t$ represents an individual term in the feedback documents.

Finally, the top-$m$ terms are interpolated with the original query using a weight $\lambda$. 
Intuitively, terms appearing in query-relevant contexts are promoted even if rare, while spurious high-frequency terms are down-weighted. 
CEQE is fully unsupervised and integrates into existing PRF pipelines with only additional BERT encoding.

\paragraph{SQET: Supervised filtering of term candidates}

\textbf{SQET}~\cite{naseri2022ceqe} turns term selection into a learned binary decision.
Given $q$ and a candidate $t$ (with or without a short context window), a cross-encoder processes the pair and outputs a relevance probability $s(t\mid q)\in[0,1]$.
When multiple mentions are available, instance-level scores $\{s_i\}$ are combined into a term-level score using simple heuristics, such as taking the maximum score across mentions, computing a normalized weighted sum where each mention is weighted by the term frequency of $t$ in its context window, or computing a normalized rank-discounted sum that down-weights mentions appearing in lower-ranked contexts returned by the first pass (e.g., using an inverse-log discount based on $\text{rank}(\text{ctx}_i)$).
The top-ranked terms are interpolated with the original query exactly as in PRF.
Because the model observes query–term pairs during training, SQET better rejects topical but off-target terms than purely unsupervised scoring.

\paragraph{BERT-QE: Selecting chunks as expansion evidence for re-ranking}

\textbf{BERT-QE}~\cite{zheng2020bert} selects short text chunks from $\mathcal{R}_k$ and reuses them as evidence when re-ranking candidates.
Starting from a first-pass pool $\mathcal{R}_k$ (optionally refined by a cross-encoder to obtain a cleaner feedback set), it splits each feedback document into overlapping chunks $\{c_i\}$ and scores each chunk with a cross-encoder relevance function $\text{rel}(q,c_i)$.
It then keeps the top-$k_c$ chunks as an evidence set $C$.
For any candidate document $d$, BERT-QE aggregates chunk-to-document relevance scores $\text{rel}(c_i,d)$ over $c_i\in C$, where more query-relevant chunks receive larger weights (assigned by a softmax over $\text{rel}(q,c_i)$).
The final re-ranking score is computed by combining the original query--document relevance with the evidence score, giving more weight to the query-vetted chunks while still preserving the original ranking signal.
Rather than adding new tokens to $q$, BERT-QE carries forward small, query-vetted excerpts that the re-ranker can match against candidate documents, yielding robust gains with a controllable compute budget.

\paragraph{CQED: Domain-aware term selection for biomedical search}

\textbf{CQED}~\cite{Khader2023LearningTR} targets scholarly biomedical search by combining general BERT with UMLS-BERT and lightweight preprocessing.
Named entities in $q$ are first identified and protected to avoid drift.
A masking module creates masked variants of $q$ to elicit semantically close substitutes.
Each mask is filled by both encoders to form a candidate pool $C=\{t\}$ that mixes general and domain-specific vocabulary.
A pairwise learning-to-rank model then scores $t\in C$ by the improvement it brings when appended to $q$; multi-head attention fuses signals from the two encoders before a final scoring layer.
The final query appends the top-$k$ selected terms from $C$ to $q$ and uses the expanded input for retrieval.
By anchoring term proposals in biomedical knowledge while keeping a general encoder in the loop, CQED reduces off-domain expansions and has reported strong gains on TREC-COVID.

\paragraph{PQEWC: Personalizing term choices from a user’s history}

\textbf{PQEWC}~\cite{bassani2023personalized} personalizes explicit expansion using a user’s past interactions.
From clicked titles or prior reads for user $u$, it builds an embedding pool $E_u=\{\mathbf{w}^{(u)}\}$ with a ColBERT encoder.
At inference, it selects expansion terms from $E_u$ either by clustering the pool and retrieving representative tokens from query-nearest clusters, or by directly retrieving the nearest tokens to the current query vector using an ANN index.
Selected terms are weighted by their similarity to the query, and the final ranking score interpolates the original query score with the contribution of the personalized terms, controlled by $\alpha$.
This design steers ambiguous or broad queries toward a user’s long-term interests at millisecond-level overhead.

Selection-based explicit QE keeps the expansion vocabulary under control, is easy to constrain with domain ontologies or user profiles, and plays well with both sparse and neural rankers.
Unsupervised scoring (CEQE) is simple and strong; supervised filtering (SQET) improves precision when labels exist; chunk selection (BERT-QE) lets a re-ranker exploit richer contextual evidence; domain and personalization layers (CQED, PQEWC) target drift and ambiguity in practice.

\subsection{Grounding and interaction}
\label{subsec:grounding_interaction}
Throughout this subsection, unless otherwise noted, $q$ denotes the original input query and $g$ a generated expansion (e.g., a passage, rationale, pseudo-document, or clue), with $G(q)=\{g_1,\dots,g_m\}$ denoting a set of such expansions. String-level concatenation is written as $[q;g]$, and we use $q^{+}$ for the augmented query (either in text form or in vector space). A first-pass retriever returns the top-$k$ pseudo-relevant documents $R_k(q)=\{d_1,\dots,d_k\}$ with scores $s(q,d)$, and we write $\mathbf{q}$ and $\mathbf{d}$ for the corresponding embeddings when discussing dense variants.

\subsubsection{Zero-Grounding, Non-Interactive QE: LLM-Driven Single-Stage Expansion}
\label{subsec:generative_plm}



This category comprises single-stage QE methods that do not rely on iterative retrieval--generation loops and are not systematically grounded in target-corpus evidence during expansion. Some methods are prompt-only, whereas others use pretrained or fine-tuned generators; the shared property is zero-grounding and non-interactive operation. 

\paragraph{Query2Doc: Few-shot pseudo-documents}
\label{subsubsec:q2d}
\textbf{Query2Doc}~\cite{wang-etal-2023-query2doc} prompts $\mathcal{L}$ to write a short passage that would plausibly answer $q$ using $k$ in-context exemplars (e.g., $k{=}4$). 
The pseudo-document $g$ is then concatenated with $q$ to obtain $q^{+}=[q; g]$.
For sparse retrieval, repeating the original query tokens before concatenation (e.g., five repetitions) preserves term salience; for dense models, $q$ and $g$ are joined with a separator and encoded jointly or separately.
Query2Doc reliably improves BM25 on MS MARCO~\cite{DBLP:conf/nips/NguyenRSGTMD16} and TREC DL (e.g., up to $+15.2\%$ nDCG@10 on DL'20) and gives modest gains when used to augment dense training.
Trade-offs include single-shot hallucinations and non-trivial decoding latency.

\paragraph{CoT-QE: Reason-then-expand prompting}
\label{subsubsec:cot}
\textbf{CoT-QE}~\cite{jagerman2023query} elicits an explanatory chain before the answer with a zero-shot prompt of the form:
``Answer the following query: \{q\}. Give the rationale before answering.'' 
The produced rationale $g$ contains definitions, disambiguation cues, and facet hints; $[q; g]$ is issued to the retriever.
A PRF-enhanced variant (CoT/PRF) injects the top-3 BM25 passages into the prompt to bias reasoning toward corpus-specific terminology.
CoT-QE outperforms classical PRF baselines and other prompt baselines on MS MARCO~\cite{DBLP:conf/nips/NguyenRSGTMD16}/BEIR~\cite{thakur2021beir}, especially on top-heavy metrics (MRR@10, nDCG@10), and remains effective with smaller models.

\paragraph{GAR: Generation-Augmented Retrieval}

\textbf{GAR}~\cite{mao2021generation} trains a seq2seq model (BART-large) on open-domain QA pairs to emit one of three query-specific strings: an answer, a declarative sentence containing the answer, or the title of a relevant passage.

At inference, the generated string $g$ is concatenated to the original query, yielding $[q; g]$, which is sent to BM25 or a dense retriever.
Running the three generators produces three rankings that can be combined by simple rank fusion for additional recall.
GAR is attractive because it is plug-and-play and task-aligned (the model learns to verbalise what a good answer looks like), but it pays generation latency and can drift if $g$ over-commits to a specific interpretation.

\paragraph{GRF: Generative Relevance Feedback without First-Pass Dependence}
\textbf{GRF}~\cite{mackie2023generative} replaces the first-pass retrieved set $R_k(q)$ with a set of LLM-generated documents $G(q)$ and estimates a relevance model directly over these generated texts. 
Each generated document contributes term signals according to its content and relevance to the query, producing an RM3-style expansion distribution without relying on the original retrieval results.
GRF prompts for corpus-matched genres such as keywords, entities, pseudo-documents, and news-style summaries, which aligns expansions with the collection discourse and yields sizable MAP and nDCG@10 gains on Robust04, CODEC, and TREC DL.

A follow-up study applies GRF to dense and learned-sparse retrievers by encoding each generated text in $G(q)$ and aggregating it with the original query embedding via a Rocchio-style update; for learned sparse models, term weights from $G(q)$ are pruned before indexing~\cite{mackie2023generative2}. 
Query-level analyses show complementarity with classical PRF: GRF improves low-recall queries, while PRF strengthens already-recalled facets. 
Weighted reciprocal-rank fusion of GRF and PRF further boosts Recall@1000 in most cases.

\paragraph{HyDE: Hypothetical document embeddings}
\label{subsubsec:hyde}
\textbf{HyDE}~\cite{gao2023precise} prompts $\mathcal{L}$ to draft one or more short, plausible passages answering $q$ (e.g., ``write a paragraph that answers the question''), then uses their embeddings as surrogates for the missing context in zero-shot dense retrieval.
Let $G=\{g_1,\dots,g_N\}$ be the generated passages and $f(\cdot)$ the fixed dense encoder. HyDE encodes each $g_i$ and averages the resulting vectors to form a pseudo-context representation, then updates the query embedding by interpolating between $f(q)$ and this pseudo-context with a weight $\alpha\in[0,1]$.
Retrieving with the updated query embedding improves zero-shot dense baselines on heterogeneous benchmarks by bridging the representation gap when no relevance labels or in-domain training are available; the main risks are drift from fanciful generations and added encoding cost for multiple $g_i$.

\paragraph{Exp4Fuse: Rank fusion for LLM-augmented sparse retrieval}
\label{subsubsec:exp4fuse}
\textbf{Exp4Fuse}~\cite{liu-zhang-2025-exp4fuse} prompts an LLM in a zero-shot setting to generate a single hypothetical document $r_q$.
To balance lexical evidence, it forms an expanded query by repeating the original query $q_o$ a fixed number of times (e.g., 5) and concatenating it with $r_q$.
Using the same sparse retriever (e.g., SPLADE++\cite{formal2022distillation}), it retrieves two ranked lists: one from $q_o$ and one from the expanded query.
The two lists are then combined with a modified reciprocal-rank fusion: for each document, it sums the usual rank-based contributions $1/(k+r)$ from both lists (with $k=60$ to reduce outlier effects), and applies a small bonus to documents that appear in both lists (based on the number of list appearances $n\in\{1,2\}$).
This indirect QE boosts nDCG@10 and Recall@1000 for learned sparse models on MS~MARCO~\cite{DBLP:conf/nips/NguyenRSGTMD16} and BEIR~\cite{thakur2021beir} with low overhead; regressions on already-strong queries suggest invoking it selectively using simple query-quality predictors.

\paragraph{Contextual clue sampling and fusion}

\textbf{Contextual clue sampling}~\cite{liu2022query} targets breadth and control. 
A generator produces many short, answer-oriented "clues" for $q$ using stochastic decoding. 
The candidate clues are clustered (e.g., by edit distance), keeping one representative per cluster to reduce redundancy. 
Each retained clue $c_i$ forms an expanded query $[q; c_i]$, producing a ranked list of candidate documents. 
The final ranking combines all these lists, giving more weight to clues that the model deems more probable.
This scheme encourages multiple reasoning paths while focusing on high-quality clues, making it particularly effective for multi-hop or under-specified questions.
\paragraph{SEAL: Generating substrings as executable keys}

\textbf{SEAL}~\cite{bevilacqua2022autoregressive} constrains generation to $n$-gram substrings that are guaranteed to occur in the target corpus.
A BART model is fine-tuned to produce such substrings, and decoding is enforced with an FM-index (a compressed full-text substring index) so that every token sequence $t$ is corpus-valid.
Each candidate substring $t$ is scored by a rarity-adjusted language-model score that balances "query plausibility" (from the BART model) and "corpus rarity" (from the FM-Index):
\begin{align}
S_{\mathrm{SEAL}}(t\mid q)=\log P_{\mathrm{LM}}(t\mid q)\;-\;\lambda\,\log P_{\mathrm{corpus}}(t),
\end{align}
where $P_{\mathrm{corpus}}(t)$ is the empirical prior of $t$ in the target corpus $\mathcal{R}$, estimated from its occurrence frequency (e.g., using the total count $F(t,\mathcal{R})$ normalized by the corpus token mass). The hyperparameter $\lambda$ balances query plausibility under the generator against corpus rarity.
This scoring formula biases toward informative (rarer) substrings that remain plausible under $q$.
Documents containing high-scoring substrings are retrieved via FM-index lookups, and evidence from multiple substrings is aggregated.
SEAL unifies expansion and retrieval: generated outputs are directly executable keys, avoiding hallucination and enabling GPU-free, low-latency search with strong accuracy on KILT-style benchmarks.

\paragraph{SU-RankFusion: Dual-layer prompt ensembles and lightweight rank fusion}
Prompt-only generative QE is often brittle: effectiveness can vary substantially with prompt wording and formatting.
To improve robustness under the chat-based system--user interface, ~\citet{li2026dual} propose
\textbf{SU-RankFusion}, which decouple (i) a behavioural \textit{system} prompt that enforces a stable
generation style (e.g., keyword-rich, query-focused, non-hallucinatory) from (ii) diversified \textit{user} prompts that
induce complementary expansion views (synonyms, facets, contexts, intent clarifications).
Each expansion is issued to a standard retriever (e.g., BM25) to obtain a ranked list, and multiple lists are then
aggregated using lightweight {SU-RankFusion} schemes.
This design targets the {zero-grounding, non-interactive} regime: it relies primarily on the LLM's prior knowledge
(without retrieval--generation iterations), while mitigating prompt sensitivity via ensembling and stabilizing downstream
ranking via list fusion.

\paragraph{Two-LLM QE: Heterogeneous expansion and refinement}
Most LLM-based QE methods rely on a single generator, which may limit the diversity and robustness of the produced expansions. To address this, ~\citet{li2026automatic} propose \textbf{Two-LLM QE}, a training-free multi-LLM expansion framework built on top of automatically pseudo-labelled domain-matched in-context exemplars. Two heterogeneous LLMs first generate complementary expansions for the same query using the same cluster-selected demonstrations, and an additional refinement LLM then consolidates them into one coherent and less noisy expansion. 
This design performs query-level fusion before retrieval.
Empirical results show that the refinement variant consistently outperforms both single-LLM expansion and direct concatenation, suggesting that controlled cross-LLM refinement is an effective way to improve the stability and effectiveness of QE.


\subsubsection{Grounding-Only, Non-Interactive QE: Corpus-Evidence Anchored Single-Pass Expansion}
\label{subsec:prompt_only}

This category covers retrieval-conditioned expansion methods in which $\mathcal{L}$ is explicitly grounded in the target collection during a single retrieve--generate--requery pass.  
Grounding can involve:  
(i) conditioning on pseudo-relevant passages $R_k(q)$;  
(ii) constraining generation to corpus-valid substrings or entities; or  
(iii) applying selection/calibration based on first-pass retrieval scores.  
Compared with Zero-Grounding QE, these methods access corpus evidence; unlike interactive QE, they do not loop over multiple passes.
Beyond the global notation, we introduce $P_{\mathrm{gen}}(g \mid q, R_k(q))$ as the generation probability given corpus context, and $\lambda \in [0,1]$ as the interpolation parameter for fusion between retrieval runs.


\paragraph{MILL: Mutual verification with corpus evidence}
\label{subsubsec:mill}
\textbf{MILL}~\cite{jia2024mill} constructs two complementary contexts and keeps only what they agree on.
Stage~1 generates multiple explanations by a query–query–document style prompt (decompose $q$ into sub-queries and write short passages), and in parallel retrieves pseudo-relevant documents with BM25.
Stage~2 performs mutual verification: generated texts are scored by their semantic consistency with retrieved documents (cosine between encoder representations), and PRF documents are rescored by their agreement with the generated explanations.
Only top-agreement items from both sides are retained, concatenated to $q$, and reissued.
Across TREC-DL'19/20 and BEIR~\cite{thakur2021beir}, MILL improves NDCG@1000, MRR@1000, and Recall@1000, with largest gains in specialized domains, indicating that agreement filtering curbs hallucination while preserving useful diversity.


\paragraph{AGR: Analyze--Generate--Refine}
\label{subsubsec:agr}
\textbf{AGR}~\cite{chen2024analyze} structures prompting into three stages. 
It first analyzes the input query $q$ by extracting key phrases and forming a brief statement of the information need. 
Next, it generates several answer-oriented candidates and, for each candidate, performs a light retrieval step to obtain a small set of supporting passages; the language model $\mathcal{L}$ then enriches each candidate using this contextual evidence. 
Finally, AGR conducts a self-review over all enriched candidates to remove off-topic or redundant content and produces a compact expansion $g$.

The resulting expanded query $[q; g]$ improves zero-shot retrieval and end-to-end OpenQA exact match across NQ, TriviaQA, WebQ, and CuratedTREC, demonstrating that staged quality control can reduce error propagation without requiring supervision.

\paragraph{EAR: Expand, rerank, and retrieve}

\textbf{EAR}~\cite{chuang2023expand} decouples generation from selection.
A generator (e.g., BART-large or T0-3B) first proposes a diverse set of candidate expansions $E=\{e_1,\dots,e_N\}$.
A learned reranker then predicts the utility of each candidate.
Two variants are common: a retrieval-independent reranker that scores pairs $(q,e)$, and a retrieval-dependent reranker that also conditions on a top passage $(p_1)$, scoring triples $(q,e,p_1)$.
With a pairwise loss, the reranker learns to prefer expansions that raise the rank of the gold passage under BM25.
The best candidate $\hat e$ is concatenated with the query and used for the final retrieval.
By filtering the generator through a PLM-based selector, EAR preserves diversity while curbing noisy $e_i$, delivering consistent gains over single-shot generators.

\paragraph{GenPRF: Generative pseudo-relevance feedback}

\textbf{GenPRF}~\cite{wang2023generative} marries classical RM3 with neural reformulation.
From the top-ranked texts of a first pass, short passages are selected (e.g., FirstP/TopP/MaxP) and fed to a seq2seq model (T5 or FLAN-T5) to produce a corpus-aware reformulation $q_r$.
Rather than replacing RM3, GenPRF mixes the two query models.
Let $p(w\mid\theta_{\mathrm{RM3}})$ be RM3’s term distribution and $p(w\mid\theta_{\mathrm{Gen}})$ the term distribution estimated from $q_r$ (e.g., normalised token weights).
GenPRF interpolates
\begin{align}
p(w\mid\theta^{\dagger}) = k_{\mathrm{RM3}}\,p(w\mid\theta_{\mathrm{RM3}})+k_{\mathrm{Gen}}\,p(w\mid\theta_{\mathrm{Gen}}),
\qquad k_{\mathrm{RM3}}+k_{\mathrm{Gen}}=1,
\end{align}
and uses $p(\cdot\mid\theta^{\dagger})$ in the second pass.
This hybrid retains RM3’s high-recall term statistics while injecting semantically richer paraphrases from the generator, yielding consistent MAP and nDCG gains across ad hoc collections.

\paragraph{CSQE: Corpus-Steered query expansion}
\textbf{CSQE}~\cite{lei2024corpus} balances LLM world knowledge with collection specificity by combining two sources of expansion text: sentences extracted from $R_k(q)$ that an LLM judges as query-relevant, and a hypothetical document generated without any corpus input. Concatenating both with $q$ improves robustness on TREC DL and BEIR~\cite{thakur2021beir}, and it is particularly effective on NovelEval, where the target knowledge is absent from LLM pretraining.

\paragraph{MUGI: Multi-text generation integration for query expansion}
\textbf{MUGI}~\cite{zhang2024exploring} prompts an LLM in a zero-shot setting to generate multiple pseudo-reference documents $R=\{r_1,\dots,r_n\}$ (e.g., $n{=}5$), and integrates them with the original query $q$ to enrich both sparse and dense retrieval without additional training.
For sparse retrieval (BM25), the pseudo-references are appended to the query, while the original query is repeated a length-normalized number of times to balance their influence, given BM25’s sensitivity to term frequency and document length (details in the appendix).
For dense retrieval, each pair $[q, r_i]$ is encoded independently and then averaged:
\begin{align}
e_q = \frac{1}{n} \sum_{i=1}^n f\big([\, q, r_i \,]\big),
\end{align}
yielding a query embedding enriched with multiple contextualized references.
To reduce noise from imperfect generations, MUGI further applies a light calibration that shifts the query embedding toward signals with high agreement (the generated references and top-$K$ retrieved documents) and away from low-ranked negatives.
This unified approach improves nDCG@10 and Recall@1000 across sparse and dense models on TREC DL and BEIR~\cite{thakur2021beir}, enabling smaller LLMs to match or surpass larger models.

\paragraph{PromptPRF: a feature-based PRF framework}
\textbf{PromptPRF}~\cite{li2025pseudo} extracts structured features from $R_k(q)$ using an LLM, such as keywords, entities, focused summaries, and chain-of-thought-derived terms, and encodes them once as an augmented query representation for the second pass. Because these features are grounded in PRDs and can be produced offline, PromptPRF reduces online LLM cost while outperforming term-statistics PRF baselines.

\paragraph{FGQE: Fair Generative Query Expansion} 
\textbf{FGQE}~\cite{jaenich2025fair}measures exposure disparities in the first-pass ranking using Average Weighted Rank Fairness and then conditions the LLM to generate entities, keywords, or pseudo-documents that target underexposed groups. The resulting expansions improve fairness with minimal loss in nDCG, illustrating that feedback signals can optimize objectives beyond relevance.

Retrieval-conditioned QE covers methods that guide expansion using first-pass collection evidence. 
Unlike prompt-only generation, these approaches incorporate retrieval signals to ground expansion in the target corpus, closing the loop within a single retrieve–expand–requery pass. 
Typical designs condition on pseudo-relevant passages $R_k(q)$ or adjust expansion impact using retrieval scores, which improves controllability and robustness when unguided generation might drift from collection-specific terminology.

\subsubsection{Grounding-Aware Interactive QE: Multi-Round Retrieve-Expand Loops}

This category covers \textit{iterative} query expansion methods that interleave retrieval and generation over $T > 1$ rounds, with at least some expansions grounded in documents retrieved in earlier stages.  
This closed-loop setting enables progressive query refinement, going beyond the single-pass grounding of the previous category.
Building on the global notation, let the initial query be $q^{(1)} = q$.  
At iteration $t$ ($1 \le t \le T$), the model produces an expansion $g^{(t)}$ (from prior knowledge or conditioned on $R_k(q^{(t)})$) and forms:
\begin{align}
q^{(t)}_{\mathrm{exp}} = [\, q^{(t)}; g^{(t)} \,].
\end{align}
Retrieved sets $R_k(q^{(t)})$ may be combined, filtered, or re-used in later iterations, and stopping criteria may depend on retrieval budgets or observed gains.

\paragraph{InteR: a novel framework that facilitates information refinement through synergy between RMs and LLMs}
\textbf{InteR}~\cite{feng2024synergistic} alternates expansion and retrieval. An LLM first produces expansion text to pull in an initial set; those passages are then summarized into a knowledge collection $S$ that is interleaved with $q$ to requery. The method supports hybrid sparse–dense stacks and shows strong results on large-scale and low-resource search.

\paragraph{ProQE: progressive query expansion} 
When retrieval calls are expensive, \textbf{ProQE}~\cite{rashid2024progressive} adopts a progressive loop that retrieves only one document per iteration, lets an LLM judge and extract terms, updates the query, and repeats until a budget is met. A final chain-of-thought expansion is added before a single high-recall sweep, delivering higher effectiveness than single-shot PRF or zero-shot LLM QE under tight budgets.

\paragraph{LameR: the Language language model as Retriever} 
\textbf{LameR} ~\cite{shen-etal-2024-retrieval}performs a small first-pass retrieval, then prompts the LLM to produce multiple likely answers conditioned on those few passages, and finally issues an answer-augmented query for the second pass. This answer-first strategy improves recall without retriever fine-tuning and sets competitive zero-shot baselines on DL’19/DL’20 and BEIR~\cite{thakur2021beir}.

\paragraph{ThinkQE: Iterative thinking with corpus interaction}
\label{subsubsec:thinkqe}
\textbf{ThinkQE}~\cite{lei-etal-2025-thinkqe} alternates explicit reasoning and retrieval.
At round $t$, the model lists interpretations and related concepts, then writes a short expansion $g_t$; the expanded query $[q; g_1; \dots; g_t]$ is issued, and only new top-$k$ documents are fed back as context for the next round to encourage coverage.
After a small number of rounds, all expansions are concatenated and searched.
On TREC DL and BRIGHT, this retrieve–expand–filter loop increases both nDCG@10 and Recall@1000 versus single-shot prompting, suggesting that iterative, diversity-seeking reasoning is an effective antidote to early bias.

Grounding-aware interactive QE covers methods that iteratively interleave retrieval and generation over multiple rounds. 
Unlike single-pass retrieval-conditioned approaches, these methods progressively refine the query by conditioning expansions on documents retrieved in earlier stages, effectively closing a multi-round loop with the corpus. 
By grounding each expansion in prior retrieval results, interactive QE improves controllability and robustness, allowing the model to correct earlier mistakes, explore under-specified aspects of the query, and converge on more precise retrieval targets.


\subsection{Learning and Alignment for QE with PLMs/LLMs}
\label{subsec:aligned_qe}

This section covers methods that explicitly align large language models or retrievers to produce query expansions that improve downstream retrieval. 
We use $q$ for a query, $g$ for an expansion text, $d^{+}$ for a gold (relevant) passage, and $D^{-}$ for a set of negatives.
Encoders map text to vectors: $f(\cdot)$ denotes the student retriever, $f_T(\cdot)$ a teacher retriever.
When needed, $\mathcal{L}$ denotes an LLM used to generate $g$.
For dense training, we write a standard contrastive (InfoNCE) objective
\begin{align}
\mathcal{L}_{\mathrm{cont}}(q)=
-\log \frac{\exp(\,\mathrm{sim}(f(q), f(d^{+}))/\tau\,)}
{\exp(\,\mathrm{sim}(f(q), f(d^{+}))/\tau\,)+\sum_{d^{-}\in D^{-}}\exp(\,\mathrm{sim}(f(q), f(d^{-}))/\tau\,)},
\end{align}
with temperature $\tau$ and cosine similarity $\mathrm{sim}(\cdot,\cdot)$.
Supervised fine-tuning (SFT) minimizes negative log-likelihood of a target expansion sequence $g$ under a generator $\pi_{\theta}$:
$\mathcal{L}_{\mathrm{SFT}}=-\log \pi_{\theta}(g\mid q)$.
Direct Preference Optimization (DPO) aligns a generator $\pi_{\theta}$ to prefer $g^{+}$ over $g^{-}$ for the same $q$:
\begin{align}
\mathcal{L}_{\mathrm{DPO}}
=-\log \sigma\!\Big(\beta\big[\log\pi_{\theta}(g^{+}\!\mid\!q)-\log\pi_{\theta}(g^{-}\!\mid\!q) - \log\pi_{\mathrm{ref}}(g^{+}\!\mid\!q)+\log\pi_{\mathrm{ref}}(g^{-}\!\mid\!q)\big]\Big),
\end{align}
with inverse temperature $\beta$ and a frozen reference model $\pi_{\mathrm{ref}}$.
Parameter-efficient fine-tuning (PEFT) such as LoRA can be used in any generator or encoder below to reduce memory and time.


\subsubsection{SoftQE: distilling LLM-generated expansions into query representations}
\label{subsubsec:softqe}
\textbf{SoftQE}~\cite{pimpalkhute2024softqe} converts the benefit of LLM expansions into a cheaper retriever that no longer requires the LLM at inference.
Offline, a large language model (e.g., GPT-3.5-turbo) generates a pseudo-document $g$ for each query $q$.
A teacher dual-encoder $f_T(\cdot)$ is trained on expanded inputs $[q; g]$ with a contrastive objective, producing a strong representation $f_T([q; g])$.
A student encoder $f(\cdot)$ then learns, from only the raw query $q$, to approximate this expansion-aware representation via an MSE distillation loss:
\begin{align}
\mathcal{L}_{\mathrm{dist}}=\big\|f(q)-f_T([q; g])\big\|_2^2.
\end{align}
This distillation term is optimized jointly with the standard contrastive objective on raw queries, teaching the student to embed $q$ as if the helpful expansion were already present.
The result is an expansion-aware retriever with the latency of a standard encoder.
In practice, SoftQE yields sizable NDCG@10 gains on MS~MARCO~\cite{DBLP:conf/nips/NguyenRSGTMD16} and BEIR~\cite{thakur2021beir} while avoiding on-the-fly decoding costs; adding cross-encoder distillation further sharpens decision boundaries.

\subsubsection{RADCoT: retrieval-augmented distillation of chain-of-thought expansions}
\label{subsubsec:radcot}
\textbf{RADCoT}~\cite{lee2024radcot} compresses a reasoning-capable teacher (e.g., GPT-3) into a smaller retrieval-augmented student that can produce chain-of-thought (CoT) style expansions efficiently.
Stage one collects pairs $(q, g^{\mathrm{CoT}})$ by prompting the teacher to explain and answer $q$.
Stage two trains a Fusion-in-Decoder (FiD) student conditioned on retrieved contexts $P=\{p_1,\dots,p_K\}$, minimizing the likelihood loss
\begin{align}
\mathcal{L}_{\mathrm{RADCoT}}=-\log P_{\mathrm{FiD}}\!\big(g^{\mathrm{CoT}} \mid q, P\big).
\end{align}
At inference, the student generates a short rationale $g^{\mathrm{CoT}}_{\mathrm{stu}}$ grounded in the passages and appends it to $q$.
Because the student relies on retrieval rather than memorized world knowledge, it achieves teacher-like expansion quality with orders-of-magnitude fewer parameters; using PEFT (e.g., LoRA adapters) is sufficient for stable training.

\subsubsection{ExpandR: alternating retriever training and preference-aligned generation}
\label{subsubsec:expandr}
\textbf{ExpandR}~\cite{yao2025expandr} jointly optimizes a retriever and an LLM so that expansions are informative and useful to the ranker.
It alternates between updating the retriever given the current expansions and updating the generator given the current retriever.

Given $q$ and an LLM expansion $g$, ExpandR forms an augmented query embedding by combining the original query embedding $f(q)$ with the embedding of the expanded input $f([q;g])$ (a simple average is used by default, and gating is optional for adaptive weighting). The retriever is then trained with $\mathcal{L}_{\mathrm{cont}}(q)$ to exploit the additional semantic signal from $g$.

In the second step, the retriever is frozen and the LLM is updated with DPO using preference pairs $(g^{+}, g^{-})$. The preference signal combines self-consistency with the LLM’s own answer conditioned on $d^{+}$ and retrieval utility, such as the rank of $d^{+}$ when searching with $[q; g]$. This update nudges the generator toward expansions that improve ranking without training a separate reward model.

By repeating these steps, the retriever becomes expansion-aware while the generator becomes retriever-aware, producing generalizable gains across MS~MARCO~\cite{DBLP:conf/nips/NguyenRSGTMD16} and BEIR~\cite{thakur2021beir}, especially on collections with large lexical--semantic gaps.





\subsubsection{Aligned Query Expansion: SFT and DPO for retrieval-oriented generation}
\label{subsubsec:aqe}
\textbf{AQE}~\cite{yang2025aligned} starts from a base LLM and explicitly reshapes its output distribution toward retrieval-helpful expansions.
For each $q$, the base model proposes a pool of candidates $\{g_i\}_{i=1}^{n}$; each candidate is scored by how highly $d^{+}$ ranks when searching with $[q; g_i]$.
AQE supports two alignment regimes.

\vspace{0.5em} 
\begin{itemize}
    \setlength{\leftmargin}{1.5em}
    \setlength{\labelwidth}{0.5em}
    \setlength{\labelsep}{0.5em}
    \item[\textemdash] \textit{RSFT.}
    Keep only top candidates as targets and run supervised fine-tuning with the token-level loss $\mathcal{L}_{\mathrm{SFT}}$. This anchors the model on high-utility expansions.

    \item[\textemdash] \textit{DPO.}
    Form preference pairs $(g^{+}, g^{-})$ from the top and bottom of the pool and optimize $\mathcal{L}_{\mathrm{DPO}}$ against a frozen reference. DPO is a light-weight alternative to RLHF that avoids training a separate reward model while still encoding utility-aware preferences.
\end{itemize}

At inference, AQE generates a single short $g$ via greedy decoding and concatenates it to $q$, eliminating generate–rank–filter loops.
On MS~MARCO~\cite{DBLP:conf/nips/NguyenRSGTMD16} and NQ~\cite{kwiatkowski2019natural}, both RSFT and DPO variants improve NDCG@10 and Recall@1000; combining RSFT (to stabilize behavior) followed by DPO (to sharpen preferences) is often strongest.
PEFT substantially reduces fine-tuning cost without hurting effectiveness.

Learning and alignment based QE adapts pretrained language models or retrievers to produce query expansions that improve downstream retrieval. 
Rather than relying on generic or zero-shot expansions, these methods use supervision, distillation, or self-alignment to guide which expansions are most effective. 
This alignment improves precision, robustness, and efficiency, allowing PLMs or LLMs to generate expansion-aware representations without depending on unrestricted generation.


\subsection{KG-augmented Query Expansion}
\label{subsec:KG-augmented}
We consider a knowledge graph $G=(V,E,R)$ with entity nodes $V$, relation types $R$, and edges $E\subseteq V\times R\times V$. Given a query $q$, let $\mathcal{E}_q$ denote entities extracted from $q$ (via LLM). These entities are further linked to nodes in $G$ via NER + entity linking to ensure alignment with KG semantics. $\mathcal{N}_h(v)$ denotes the $h$-hop neighborhood of $v\in V$. We write $\text{text}(v)$ for a textual field of an entity (label, aliases, or description), and use $\mathrm{concat}(\cdot)$ to denote textual concatenation. Unless stated otherwise, the final retrieval is performed with a sparse or dense retriever scoring $s(d\mid q^{+})$ on an expanded query $q^{+}$.

It should be noted that KG-augmented QE here is not independent of the "grounding" framework in Sec.\ref{subsec:grounding_interaction}: Sec.\ref{subsec:grounding_interaction} defines "grounding" as "expansion signals relying on real-time evidence from the current retrieval corpus", while this section takes external KGs as a new grounding source, forming a complementation of "corpus grounding (Sec.\ref{subsec:grounding_interaction}) + KG grounding (this section)". Corpus grounding ensures alignment with the target corpus, while KG grounding supplements structured semantics missing in the corpus, jointly addressing complex retrieval needs.

After introducing the basic concepts of knowledge graphs, we note that the key factor in KG-augmented query expansion lies in how the extracted entities are utilized. Existing approaches can be broadly divided into two categories: (i) leverages knowledge graph retrieval to incorporate related entities and relations. (ii) links the extracted entities to external corpora or text chunks to obtain additional expansion information.

\subsubsection{Entity-based Expansion via Knowledge Graph Retrieval}
In this category of methods, entities are first extracted from the query, and then their neighboring nodes or relations in the knowledge graph are retrieved to enrich the query representation. The expanded query can then be expressed as:
\begin{align}
q^+ = \big[q \; ; \; \bigcup_{e \in \mathcal{E}_q} f(\mathcal{N}(v), \mathrm{R}(v))\big],
\end{align}
where $f(\cdot)$ is a function that selects or weights expansion terms from neighboring entities and relations, and $[q; \cdot]$ denotes concatenation or fusion with the original query.
In the following, we present specific approaches for this category.

\paragraph{Entity-aware text injection (KGQE)}
\textbf{KGQE}~\cite{perna2025knowledge} injects compact KG facts directly into the query text to improve disambiguation and coverage. First, entities are extracted from the query, then linked to nodes in the target KG to ensure alignment with KG semantics. Then short textual snippets for each $e\in\mathcal{E}_q$ (e.g., label, type, canonical name) are selected and concatenated:
\begin{align}
q^{+} \;=\; \mathrm{concat}\Big(q;\; \big[\,\text{text}(e)\,\big]_{e\in\mathcal{E}_q}\Big).
\end{align}
Optional guards include type filters (retain only domain-relevant types), length caps, and de-duplication. The enriched query is issued to the retriever, or used to prompt an LLM when a downstream generator is present. A minimal prompt template is:

\begin{verbatim}
{
  QUERY: "When is the last episode of season 8 of The Walking Dead?",
  INJECTION: {"head": "The Walking Dead", "type": "TV Series", "publisher": "AMC"},
  ENRICHED QUERY: "When is the last episode of season 8 of The Walking Dead? [SEP]
                   The Walking Dead — TV Series — AMC"
}
\end{verbatim}

This family keeps expansions knowledge-grounded and controllable while reducing ambiguity of short or entity-heavy queries.
KGQE addresses entity ambiguity (e.g., disambiguating "Apple" as a company vs. fruit) by injecting type-specific KG facts (e.g., "Apple Inc. - Tech Company - Cupertino")—a key advantage over traditional WordNet-based expansion, which fails to resolve short-query polysemy. However, KGQE struggles with low-resource domains (e.g., niche medical subfields) where KG coverage is sparse; in such cases, hybrid schemes (e.g., KGQE + BioBERT term selection \cite{Lee2019BioBERTAP}) are required to maintain effectiveness.
\paragraph{Cross-lingual KG-aware expansion (CL-KGQE)}
\textbf{CL-KGQE}~\cite{rahman2019query} targets cross-language mismatch by combining translation, KG linking, and distributional neighbors. An input query in language $\ell_s$ is translated or linked cross-lingually to entities in a target KG; expansions are assembled from three sources: (i) distributional semantic neighbors of query tokens (word embeddings), (ii) KG categories or types (e.g., DBpedia \texttt{dct: subject}), and (iii) hypernym or hyponym terms from lexical graphs. The final expansion pool is merged with the translated keywords and executed with a standard ranker (e.g., BM25). This hybrid design improves robustness when parallel data are scarce, and terminology diverges across languages.
\subsubsection{Entity-based Expansion via Hybrid KG and Document Graph}
In this approach, each query entity $e \in \mathcal{E}_q$ is first linked to a set of relevant documents or chunks containing $e$, denoted as $D(e)$. 
A document-level graph is then constructed where nodes correspond to these documents, and edges represent relationships such as citations, authorship, or co-occurrence of entities. 
Expansion terms are obtained from the graph nodes through a selection function $g(\cdot)$, and concatenated with the original query:

\begin{align}
q^+ = \big[q \; ; \; \bigcup_{e \in \mathcal{E}_q} g(D(e))\big],
\end{align}

where $q^{+}$ is the expanded query, $[q; \cdot]$ denotes concatenation, and $g(\cdot)$ extracts the most relevant terms from the document graph.

\paragraph{KG-guided generation and retrieval (KAR)}
\textbf{KAR}~\cite{xia-etal-2025-knowledge} leverages structural neighborhoods to guide expansion. After linking $\mathcal{E}_q$, retrieve an $h$-hop subgraph $\mathcal{S}_q=\bigcup_{e\in\mathcal{E}_q}\mathcal{N}_h(e)$ and align documents to nodes/edges. From $\mathcal{S}_q$, construct lightweight relational snippets (``document triples''):
\begin{align}
T_q=\big\{(d_i, r_{i,j}, d_j)\mid (v_i,r_{i,j},v_j)\in \mathcal{S}_q,\ d_i\!\leftrightarrow\!v_i,\ d_j\!\leftrightarrow\!v_j\big\}.
\end{align}
An LLM is then prompted with $(q, T_q)$ to write a short, knowledge-grounded expansion $g_q=\mathrm{LLM}(q,T_q)$, and the final expanded query is $q^{+}=\mathrm{concat}(q; g_q)$. Compared with pure text injection, KAR exploits relational context to surface salient, query-specific facets, which is particularly effective on semi-structured corpora (e.g., author–paper–venue graphs).

\paragraph{Query-specific KG construction (QSKG)}
\textbf{QSKG}~\cite{mackie2022query} builds a small, task-specific graph on the fly. A first-pass retrieval for $q$ yields $R_k(q)$; entities in these documents are linked and added as nodes, with edges created via within-document co-mentions or KG links, expanding iteratively to form $G_q$. Node salience is computed with a simple centrality or TF–IDF style score over $G_q$, prioritizing entities/relations tightly connected to the query intent; top entities and relations are verbalized into short expansions and concatenated with the query:
\begin{align}
q^{+} \;=\; \mathrm{concat}\Big(q;\; \big[\text{text}(v)\big]_{v\in \mathrm{TopK}(G_q)}\Big).
\end{align}
QSKG adapts to collection-specific vocabulary and senses, and tends to improve recall on ambiguous or multi-facet queries while limiting drift through graph-based salience.

KG-augmented QE improves disambiguation and coverage but introduces practical choices about scope and control. In practice, injected text should be kept short (e.g., entity labels, types, and one or two discriminative facts), and drift is best controlled by relying on entity-linking confidence together with simple type filters. For KAR/QSKG-style methods, it is also important to cap the graph radius and select only a small number of high-yield nodes based on salience. In cross-lingual settings, KG categories and lexical graphs that transfer reliably across languages are usually more robust than language-specific surface forms. These strategies preserve efficiency and mitigate noise while delivering consistent gains on entity-centric and long-tail queries.

\section{Application Domains and Use Cases}
\label{sec:applications}

This section reviews where query expansion (QE) matters in practice and how traditional methods and PLM/LLM-based approaches are instantiated across domains. We focus on the mismatch patterns QE must bridge, such as short queries versus rich documents, lay versus technical vocabulary, and natural language versus code symbols. We also summarize what evidence and signals are typically available in each domain, including logs, taxonomies, ontologies, bilingual resources, dialogue history, and repositories. Finally, we highlight the deployment constraints that shape feasible instantiations, including latency budgets, the risk of drift or hallucination, and validity constraints.

Table~\ref{tab:master_domain_matrix} provides a compact overview of these application domains, highlighting the dominant mismatch/constraints and the evidence sources typically available for QE.

\begin{table*}[t]
\centering
\small
\setlength{\tabcolsep}{4pt}
\renewcommand{\arraystretch}{1.15}
\caption{Domains, primary mismatch/constraints, and evidence sources for QE.}
\label{tab:master_domain_matrix}
\begin{tabularx}{\textwidth}{p{0.16\textwidth} p{0.4\textwidth} X}
\toprule
\textbf{Domain} &
\textbf{Primary mismatch / constraint} &
\textbf{Evidence / knowledge sources} \\
\midrule
Web search &
Short, diverse queries; extreme QPS; tight latency &
Query logs, clicks, sessions~\cite{cui2003query,kunpeng2009new} \\
\addlinespace[2pt]
E-commerce search &
Colloquial intent $\rightarrow$ structured catalog; long-tail and SKU drift &
Catalog attributes/taxonomy~\cite{homoceanu2014querying,kang2024improving}; behavior logs~\cite{mandal2019query} \\
\addlinespace[2pt]
Biomedical IR &
Lay--technical vocabulary gap; high factual risk &
Ontologies (UMLS/MeSH)~\cite{aronson1997query}; PubMed corpora; domain PLMs~\cite{kelly2021enhancing,khader2022contextual} \\
\addlinespace[2pt]
Cross-lingual / multilingual search &
Translation ambiguity; low-resource coverage &
MT/dictionaries; cross-lingual embeddings; bilingual resources~\cite{rajaei2024enhancing} \\
\addlinespace[2pt]
ODQA / Knowledge-intensive retrieval &
Question $\rightarrow$ evidence retrieval (recall + precision) &
Retrieved passages; generated contexts~\cite{mao2021generation} \\
\addlinespace[2pt]
RAG &
Evidence pollution directly affects generation faithfulness &
Top-$k$ retrieved passages; QA exemplars; KG facts~\cite{wu2025kg} \\
\addlinespace[2pt]
Conversational search &
Coreference/ellipsis; evolving intent across turns &
Dialogue history; turn selection signals~\cite{mo2023learning,voskarides2020query} \\
\addlinespace[2pt]
Code search &
NL--code mismatch; syntax validity and identifier grounding &
APIs/docs; repositories; issues~\cite{nie2016query,li2022generation} \\
\bottomrule
\end{tabularx}
\end{table*}


\subsection{Web Search Engines}

\begin{figure*}[htbp]
  \centering
  \includegraphics[width=0.8\textwidth,trim=10 10 10 10,clip]{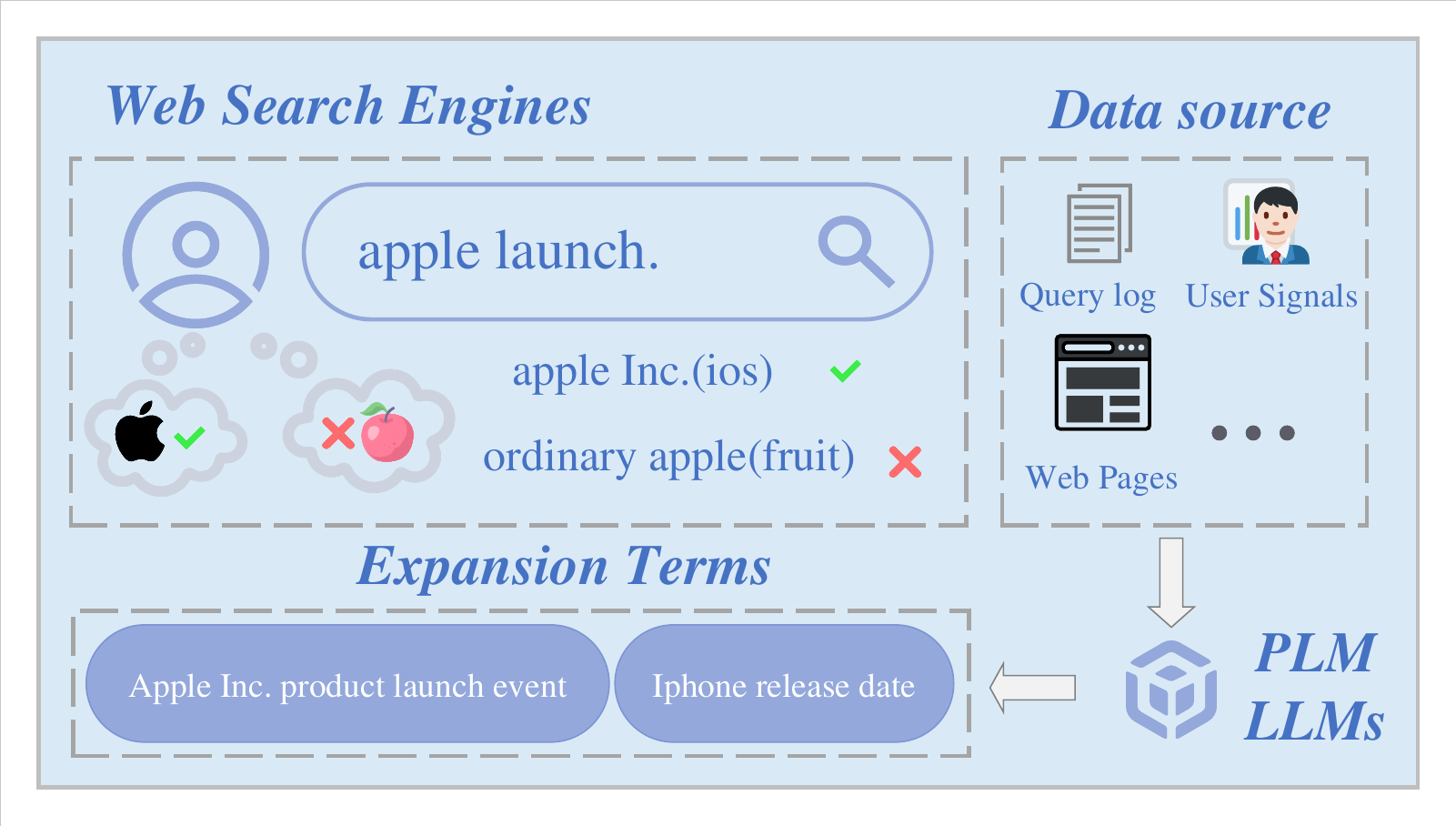}
  \caption{QE in Web Search: Pipeline and Insertion Points}
  \label{fig:app_web}
\end{figure*}

Fig.~\ref{fig:app_web} places QE in a web-scale pipeline. In this setting, QE is largely a high-throughput engineering problem: systems must improve recall for short, underspecified queries without violating tight latency constraints.

\vspace{0.5em} 
\begin{itemize}
    \setlength{\leftmargin}{1.5em}
    \setlength{\labelwidth}{0.5em}
    \setlength{\labelsep}{0.5em}
    
    \item[\textemdash] \textit{From lexical rewriting to semantic interpretation.}
    Commercial web search has long relied on query rewriting and synonym expansion to cope with short, diverse user queries. Early pipelines (e.g., spelling correction \citep{zhou2025training}, query substitution \citep{jones2006generating}) increased recall but often lacked robust intent understanding, leading to noisy expansions. 
    Modern web search increasingly incorporates neural semantic models and, in some settings, foundation models to better interpret short and ambiguous queries
    ~\cite{wang-etal-2023-query2doc,achiam2023gpt}, improving recall--precision balance compared to purely lexical heuristics.

    \item[\textemdash] \textit{Behavioral signals at scale.}
    Large query logs provide rich signals to mine reformulations and near-synonyms from user behavior~\cite{cui2003query}. Beyond pairwise mining, industrial systems aggregate candidates from behavior, taxonomy, and KBs, followed by ML-based filtering to retain expansions that improve precision. Research also leverages multi-session clicks, dwell time, and query co-occurrence to select expansion terms aligned with user intent~\cite{kunpeng2009new,cui2003query}.
    
    Practical note: Due to QPS constraints, always-on generative QE is uncommon; heavier expansion is typically gated to tail/ambiguous queries and fused conservatively.

    \item[\textemdash] \textit{Event-centric expansion with PLMs for time-sensitive intents.}
    To better handle breaking-news scenarios, Event-Centric Query Expansion (EQE)~\cite{zhang2023event} extends a query with the most salient ongoing event via a four-stage pipeline: event collection, event reformulation (leveraging PLMs such as BART and mT5 with prompt tuning and contrastive learning), semantic retrieval (a fine-tuned RoBERTa-based dual-tower model with two-stage contrastive training), and lightweight online ranking. This system has been deployed in Tencent QQ Browser Search to serve hundreds of millions of users.
\end{itemize}

\subsection{E-commerce Search}
\begin{figure*}[htbp]
  \centering
  \includegraphics[width=0.8\textwidth,trim=10 10 10 10,clip]{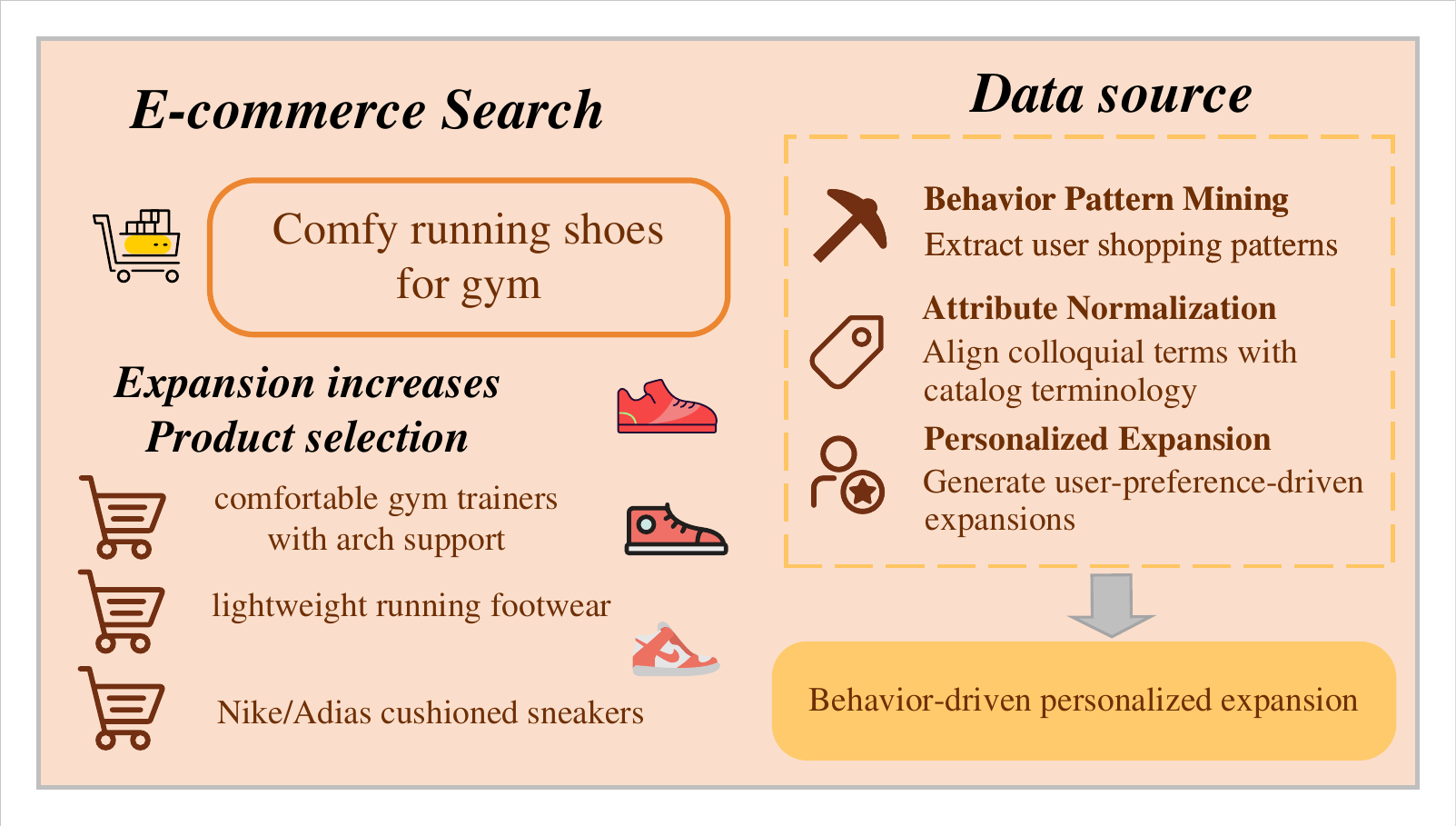}
  \caption{QE in E-commerce Search: Catalog-Aware Pipeline and Insertion Points}
  \label{fig:app_ecom}
\end{figure*}

Fig.~\ref{fig:app_ecom} highlights the e-commerce setting. While it shares the high-QPS constraint of web search, it introduces a distinctive mismatch: users express intent colloquially while the retrieval target is a structured catalog with attributes and inventory constraints.

\vspace{0.5em}
\begin{itemize}
    \setlength{\leftmargin}{1.5em}
    \setlength{\labelwidth}{0.5em}
    \setlength{\labelsep}{0.5em}

\item[\textemdash] \textit{Catalog mismatch and normalization.}
User queries are short and colloquial, while product catalogs are verbose and taxonomy-driven; QE aligns lay intent with normalized attributes~\cite{homoceanu2014querying}. Synonym mining from logs plus ML filtering (e.g., eBay) improves recall and engagement in production~\cite{mandal2019query}.
Practical note: expansions are often validated against schema and inventory (e.g., attribute constraints) to avoid retrieving out-of-catalog or incompatible items.

\item[\textemdash] \textit{Generative and taxonomy-aware enrichment.}
Beyond synonymy, mQE-CGAN~\cite{cakir2023modified} leverages GANs to synthesize expansions conditioned on lexical/semantic signals, outperforming strong baselines (e.g., BART/CTRL) on semantic similarity and coverage. ToTER~\cite{kang2024improving} uses topical taxonomies for silver labeling, topic distillation, and query enrichment, improving Recall/NDCG/MAP over SPLADE++~\cite{formal2022distillation}, ColBERT, PRF, and recent generative augmentation across Amazon ESCI.

\item[\textemdash] \textit{LLM rewrites at scale.}
LLM-based personalization and on-the-fly rewrites normalize attributes and surface rare variants~\cite{wang2024leveraging}. Incorporating intent signals (name/category prediction) with auxiliary losses improves long-tail rewriting in practice~\cite{zhang2022advancing}. Taobao’s BEQUE~\cite{peng2024large} aligns rewrites to retrieval utility via offline feedback and Bradley--Terry-style objectives, yielding measurable GMV/CTR gains in a 14-day A/B test, with pronounced lifts on tail and few-recall queries.

\end{itemize}


\subsection{Biomedical Information Retrieval}
\begin{figure*}[htbp]
  \centering
  \includegraphics[width=0.8\textwidth,trim=10 10 10 10,clip]{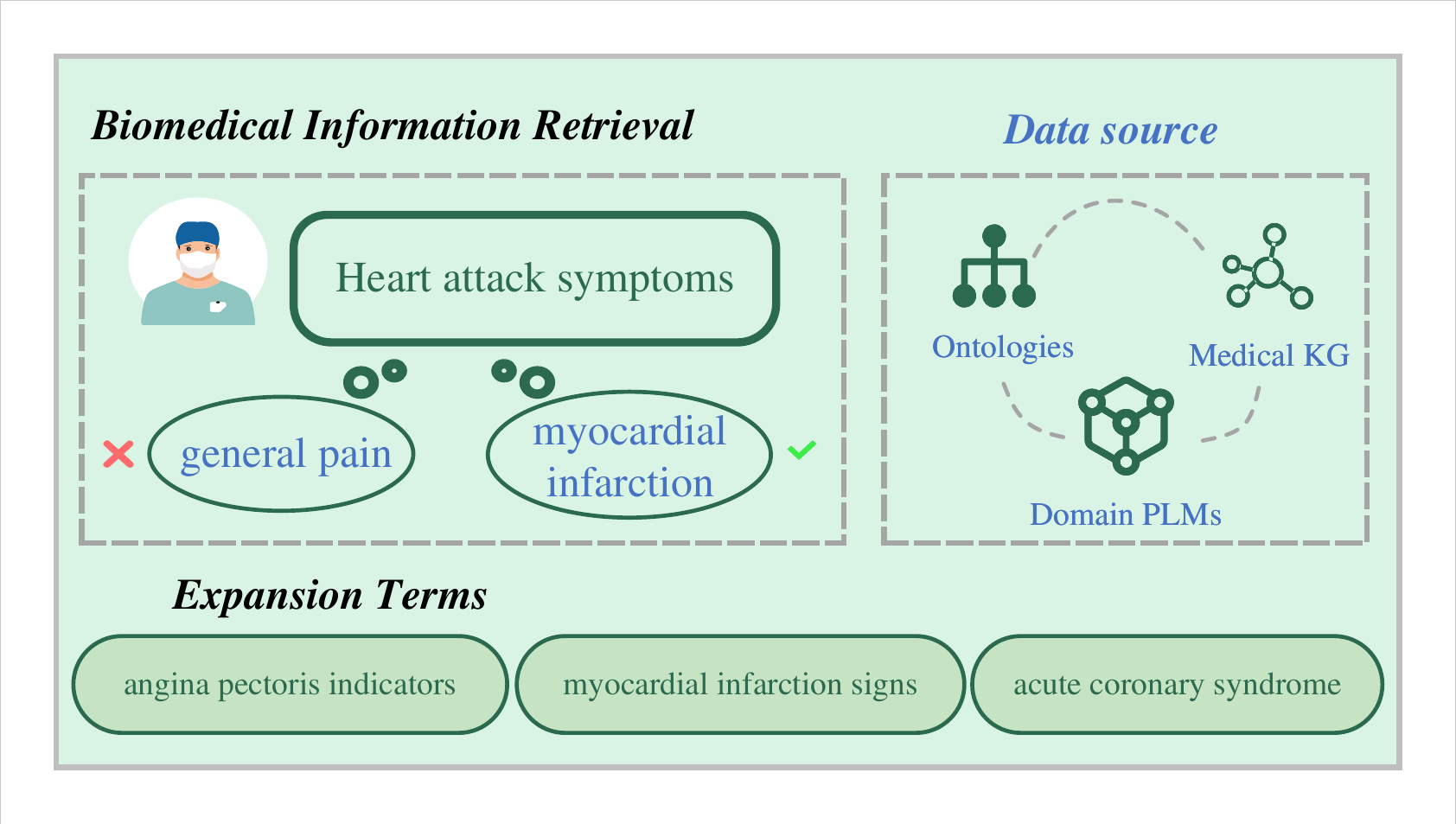}
  \caption{Biomedical IR: ontology/KG-grounded QE workflows and safety-critical controls.}
  \label{fig:app_biomed}
\end{figure*}

Biomedical IR (Fig.~\ref{fig:app_biomed}) is characterized by severe lay--technical vocabulary gaps and high factual risk. Consequently, QE is typically grounded in ontologies/KGs and domain corpora, and unconstrained generation is treated cautiously.

\vspace{0.5em}
\begin{itemize}
    \setlength{\leftmargin}{1.5em}
    \setlength{\labelwidth}{0.5em}
    \setlength{\labelsep}{0.5em}

\item[\textemdash] \textit{Specialized vocabulary and ontologies.}
Biomedical IR faces severe lay--technical vocabulary gaps (“heart attack” vs.\ “myocardial infarction”). Ontology-based QE (UMLS, MeSH) maps free text to controlled concepts and spelling variants (e.g., MetaMap + ATM), improving classic metrics and reducing ambiguity. \citet{aronson1997query} report up to +14.1\% 11-pt AP on the Hersh collection versus non-concept baselines, illustrating the benefit of structured medical knowledge.

\item[\textemdash] \textit{Domain PLMs.}
Domain-adapted PLMs (e.g., BioBERT, PubMedBERT) recommend contextual terms and capture fast-evolving terminology. \citet{kelly2021enhancing} show BioBERT-based embedding QE for rare diseases substantially improves Precision@10 (0.42 $\to$ 0.68) and nDCG (0.48 $\to$ 0.72) over BM25/MeSH, with strong gains on acronym queries. For systematic reviews, \citet{khader2022contextual} combine BioBERT signals with MeSH features in a learning-to-rank framework, further boosting P@10/MAP/Recall; hybridizing semantic and ontology signals is especially effective.

\item[\textemdash] \textit{Interactive LLM workflows.}
\citet{ateia2025bioragent} design an interactive biomedical RAG agent that prompts an LLM to emit an editable expanded query (Elasticsearch DSL), enabling experts to inspect and prune LLM-generated terms before BM25 retrieval. This strikes a balance between semantic priors and expert control.

\item[\textemdash] \textit{Precision and hallucination control.}
High-stakes settings require drift control: UMLS term reweighting with self-information can mitigate topic drift~\cite{diao2018research}. For LLMs, \citet{niumitigating} propose self-refinement guided by predictive uncertainty to detect risky entities and selectively verify them against medical KGs, improving factuality with limited overhead.

\end{itemize}


\subsection{Cross-Lingual \& Multilingual Search}

\begin{figure*}[htbp]
  \centering
  \includegraphics[width=0.8\textwidth,trim=10 10 10 10,clip]{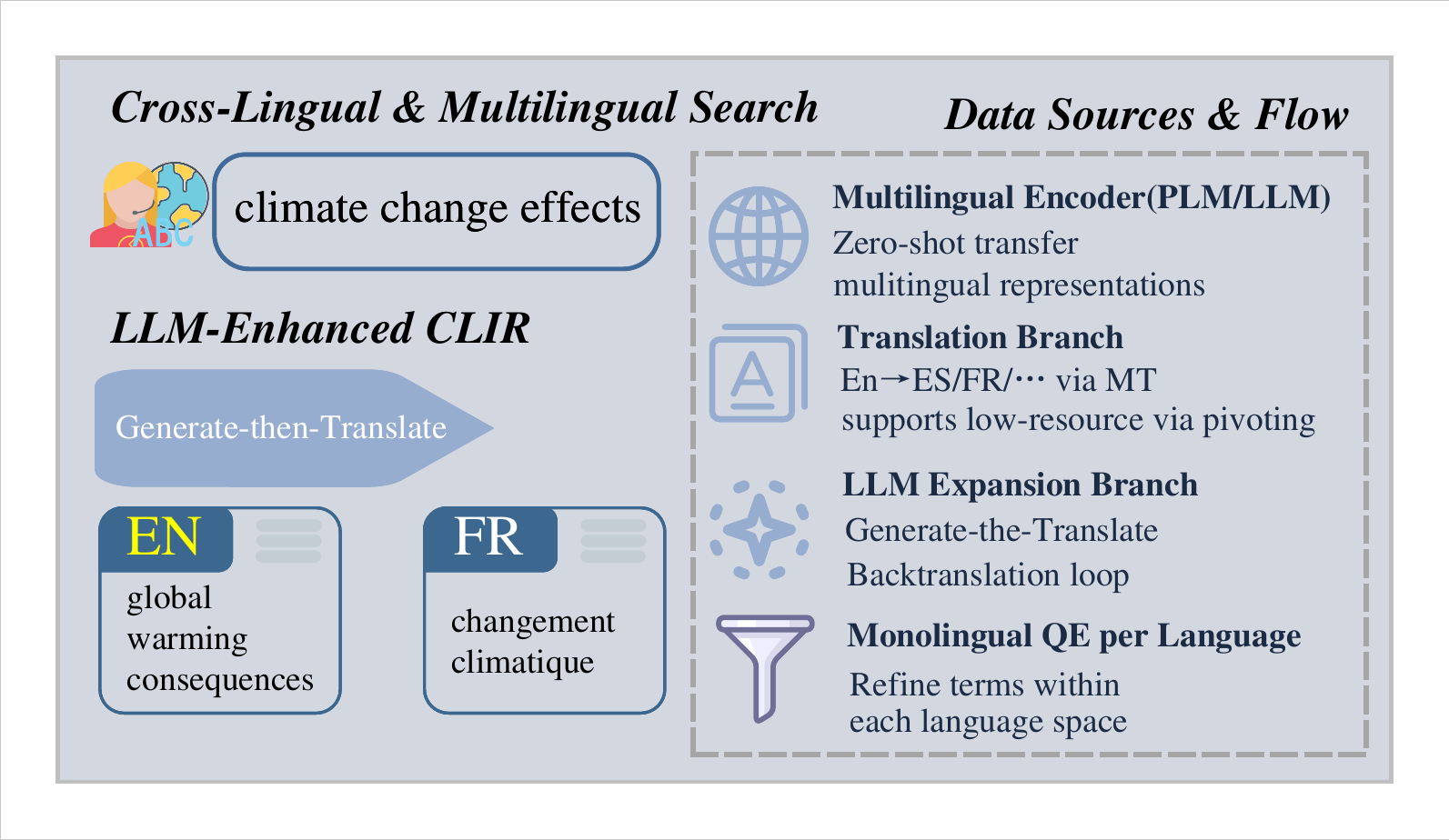}
  \caption{Cross-lingual \& multilingual search: alignment/translation-centric QE workflows.}
  \label{fig:app_clir}
\end{figure*}

Fig.~\ref{fig:app_clir} summarizes alignment/translation-centric workflows. Traditional CLIR pipelines translate queries via dictionaries/MT and then apply monolingual QE; low-resource coverage and ambiguity remain key challenges. In contrast, multilingual PLMs/LLMs enable stronger representation alignment and support generate-then-translate or backtranslation strategies.

For example, \citet{rajaei2024enhancing} reformulate queries via multi-language backtranslation and fuse ranked lists with RRF, improving MAP on classic testbeds (e.g., Robust04, GOV2). In practice, combining MT with LLM-driven expansions can increase cross-language recall while controlling drift via conservative fusion and entity-preservation heuristics.


\subsection{Open-Domain Question Answering}
\begin{figure*}[htbp]
  \centering
  \includegraphics[width=0.8\textwidth,trim=10 10 10 10,clip]{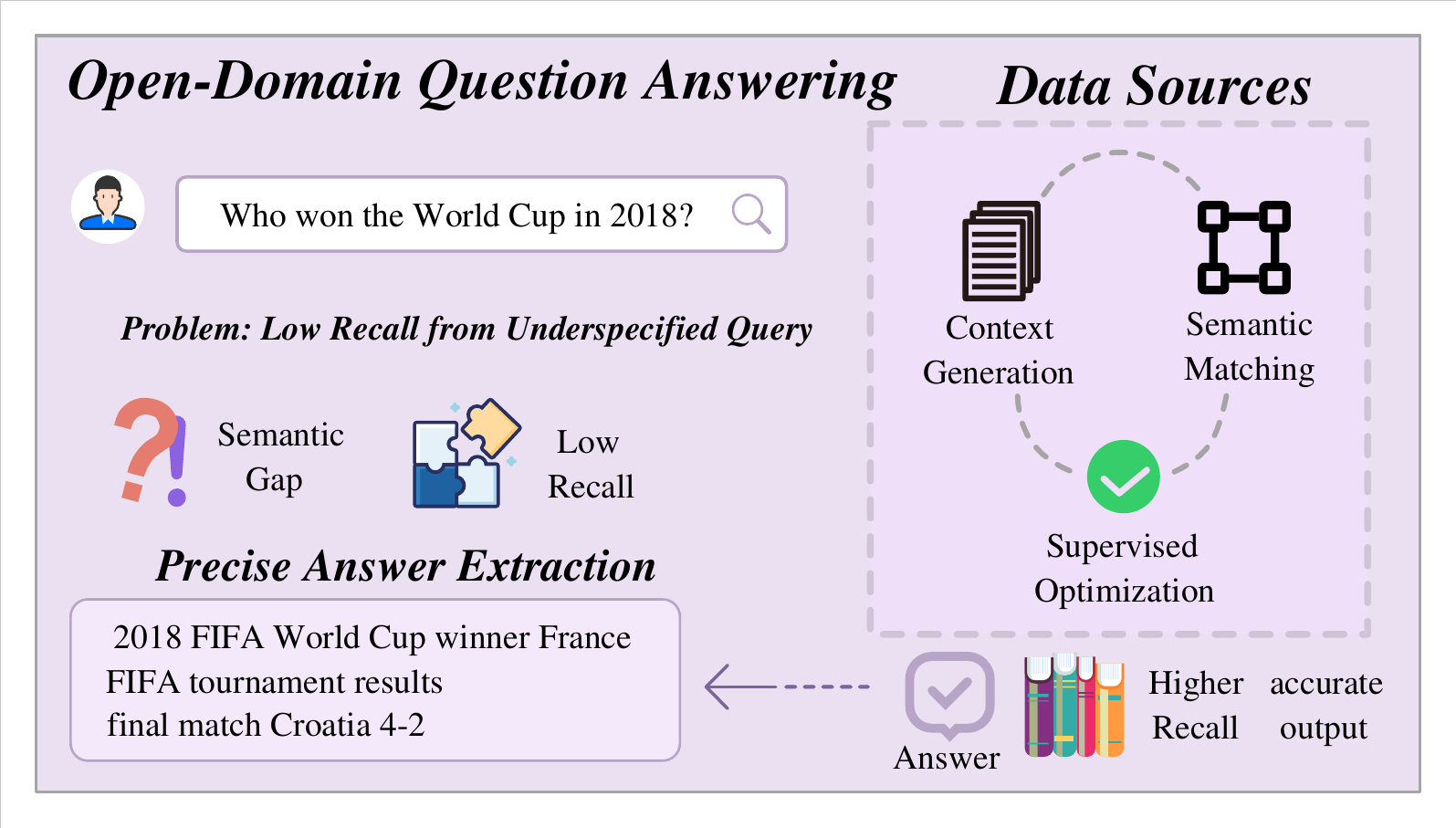}
  \caption{QE in Open-Domain QA: Evidence-Centric Retrieval Pipeline}
  \label{fig:app_odqa}
\end{figure*}

Fig.~\ref{fig:app_odqa} illustrates evidence-centric QE patterns in knowledge-intensive systems, where the objective is not only to match vocabulary but to retrieve answer-bearing evidence.

\vspace{0.5em}
\begin{itemize}
    \setlength{\leftmargin}{1.5em}
    \setlength{\labelwidth}{0.5em}
    \setlength{\labelsep}{0.5em}

\item[\textemdash] \textit{Answer-oriented expansions.}
Generation-Augmented Retrieval (GAR)~\cite{mao2021generation} enriches queries with generated contexts and combines BM25-level sparse retrieval with DPR, showing that properly scoped pseudo-text can rival dense pipelines. EAR~\cite{chuang2023expand} further couples expansion with reranking. Recent AQE~\cite{yang2025aligned} aligns expansions with downstream passage utility.

\item[\textemdash] \textit{Zero-shot and process supervision.}
LLMs can yield high-quality expansions without task-specific fine-tuning (e.g., MILL~\cite{jia2024mill}). Analyze--Generate--Refine (AGR)~\cite{chen2024analyze} decomposes needs and quality-controls the expansions. Hybrid text generation with PRF (HTGQE)~\cite{zhu2023hybrid} combines multiple LLM-generated contexts (answer/sentence/title) to improve EM on NQ/Trivia.

\item[\textemdash] \textit{Topic-aware ICL and vocabulary projection.}
TDPR~\cite{li5367307query} clusters queries to select ICL demonstrations, generates pseudo-passages, and projects embeddings onto vocabulary to produce interpretable keywords, improving R@20 and end-to-end accuracy across NQ/Trivia/WebQ/CuratedTREC.

\end{itemize}

\subsection{Retrieval-Augmented Generation (RAG)}
\begin{figure*}[htbp]
  \centering
  \includegraphics[width=0.8\textwidth,trim=10 10 10 10,clip]{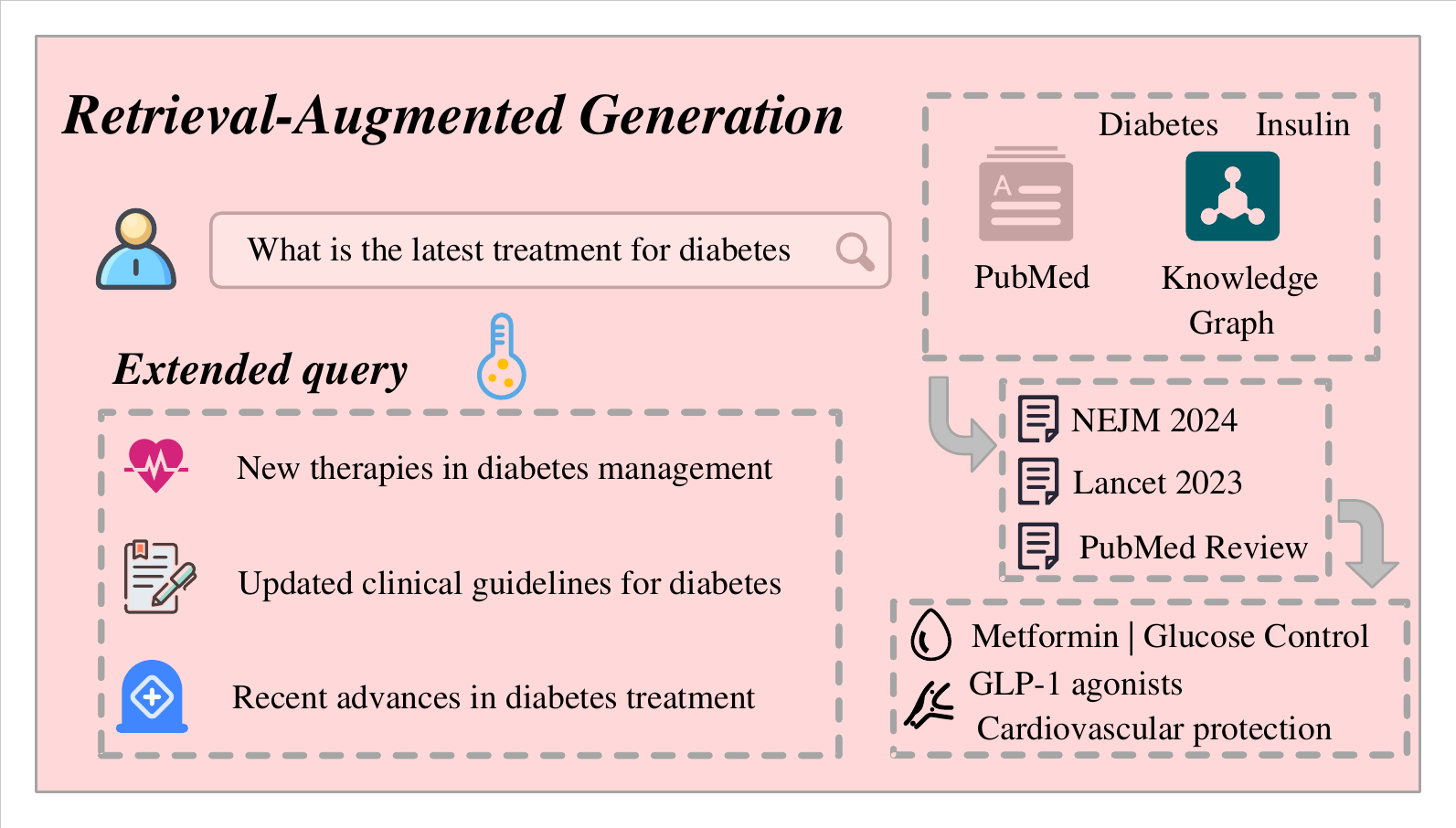}
  \caption{QE in RAG: Precision-First Evidence Retrieval for Generation}
  \label{fig:app_rag}
\end{figure*}

Fig.~\ref{fig:app_rag} shows RAG as a precision-first evidence setting. Filtering requirements are stricter because expansion noise can directly pollute the retrieved evidence used for generation.

\vspace{0.5em}
\begin{itemize}
    \setlength{\leftmargin}{1.5em}
    \setlength{\labelwidth}{0.5em}
    \setlength{\labelsep}{0.5em}

\item[\textemdash] \textit{Why QE helps RAG.}
RAG quality hinges on retrieving the right passages; vague queries increase hallucination risk~\cite{zhang2025siren,niu2024ragtruth}. Refining or expanding the query enhances retrieval coverage and answer grounding~\cite{jagerman2023query,zhang2025qe}.

\item[\textemdash] \textit{Representative strategies.}
QE-RAGpro~\cite{huquery} uses domain-tuned embeddings to fetch Q\&A exemplars for prompting LLMs to synthesize pseudo-documents, improving top-$k$ accuracy on finance datasets. QOQA~\cite{koo2024optimizing} generates multiple LLM rewrites, selects them based on query--document alignment (hybrid scores), and lifts nDCG@10 on SciFact/TREC-COVID/FiQA. KG-Infused RAG~\cite{wu2025kg} performs KG-guided spreading activation, infuses KG facts into LLM expansions, and improves across multi-hop QA benchmarks. Backtranslation-based multi-view queries fused by RRF~\cite{rajaei2024enhancing} yield consistent gains in unsupervised settings.

Practical note. Selective invocation, caching, and small/distilled models reduce latency and cost; filtering (e.g., document-aligned scoring) limits drift caused by generative variability.

\end{itemize}

\subsection{Conversational Search}
\begin{figure*}[htbp]
  \centering
  \includegraphics[width=0.8\textwidth,trim=10 10 10 10,clip]{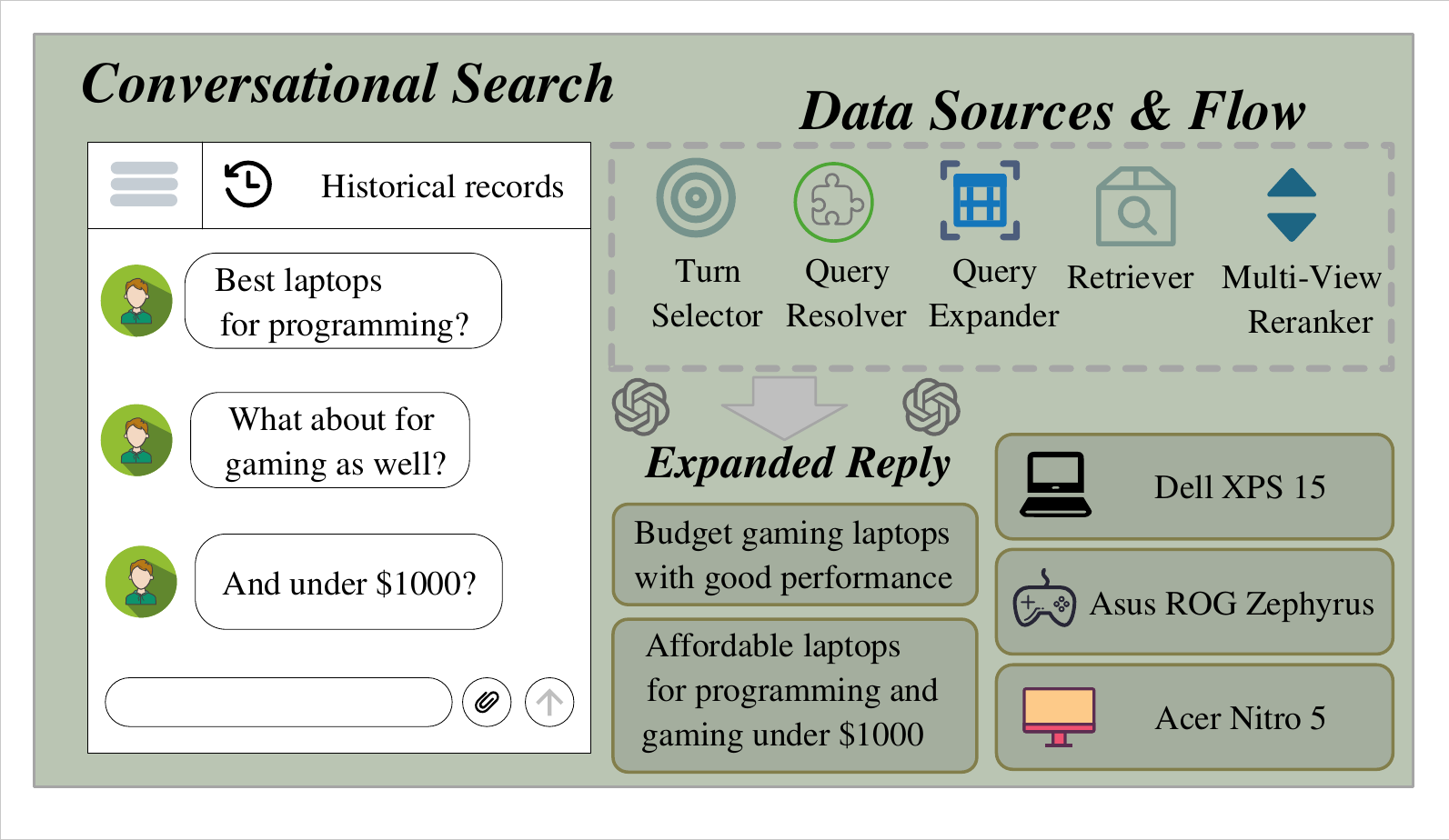}
  \caption{QE in Conversational Search: Context-Dependent Retrieval and Rewriting}
  \label{fig:app_conv}
\end{figure*}

Fig.~\ref{fig:app_conv} highlights the context-dependence of conversational QE. Systems must resolve coreference, infer omitted content, and adapt to dynamically evolving user intents---an ability traditional QE lacks, as it struggles to selectively incorporate dialogue history.

Table~\ref{tab:conv_qe} summarizes representative methods for conversational query expansion, each with distinct design focuses. Learning to select useful turns (PRL)~\cite{mo2023learning} constructs expanded queries by identifying valuable historical turns, enabling better handling of topic switches while improving MRR and NDCG metrics. MISE~\cite{kumar2020making}, which integrates Conversational Term Selection (CTS) and Multi-View Reranking (MVR), fuses multi-source views (dialogue context, relevant passages, and paraphrases) to achieve strong retrieval performance. QuReTeC~\cite{voskarides2020query} formulates term selection as a lightweight contextual classification task, effectively boosting MAP and NDCG@3 on the CAsT conversational search benchmark.

Practical note: To avoid stale context dominating the expansion process, many systems implement history selection or recency control mechanisms to filter and prioritize up-to-date dialogue information.

{\renewcommand{\arraystretch}{1.25}
\begin{table}[htbp]
\small
\centering
\caption{Representative methods for conversational query expansion.}
\label{tab:conv_qe}
\begin{tabular}{p{3.0cm} p{2.2cm} p{2.2cm} p{5.8cm}}
\toprule
\textbf{Method} & \textbf{Granularity} & \textbf{Model} & \textbf{Notable strengths} \\
\midrule
PRL~\cite{mo2023learning} & Query & BERT/ANCE & Turn selection; topic-switch robustness; higher MRR/NDCG \\
MISE (CTS + MVR)~\cite{kumar2020making} & Term / Multi-view & BERT/GPT-2 & Multi-view fusion (dialog, passage, paraphrase); strong retrieval gains \\
QuReTeC~\cite{voskarides2020query} & Term & BERT & Lightweight term selection over context; higher MAP/NDCG@3 \\
\bottomrule
\end{tabular}
\end{table}
}

\subsection{Code Search}
\begin{figure*}[htbp]
  \centering
  \includegraphics[width=0.8\textwidth,trim=10 10 10 10,clip]{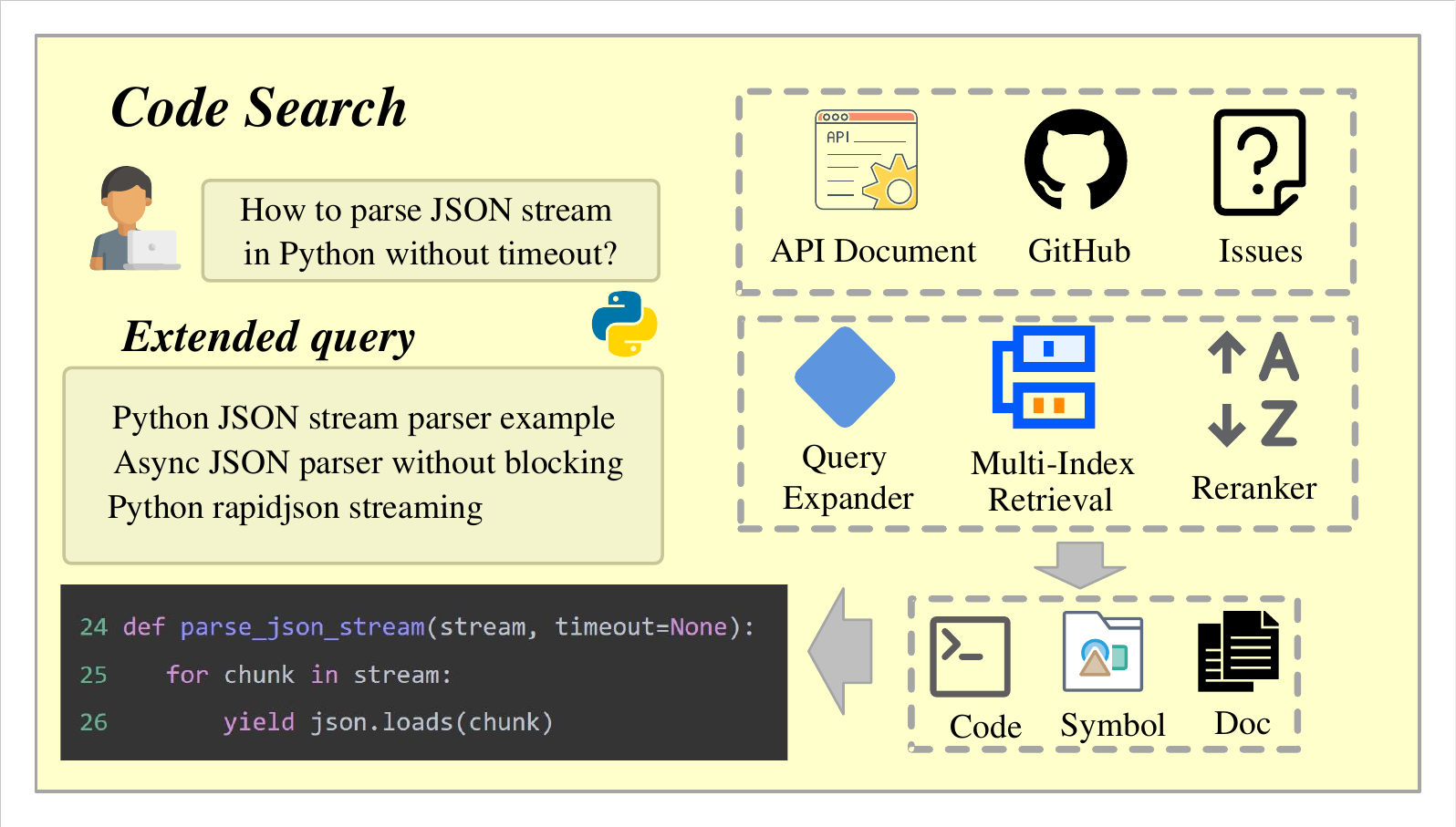}
  \caption{QE in Code Search: Symbol-Grounded Retrieval and Validity Constraints}
  \label{fig:app_code}
\end{figure*}

Fig.~\ref{fig:app_code} illustrates code search, where QE must respect symbol grounding and validity constraints. Expansions should follow syntax and stay within the repository’s identifier space.

Early work enriches queries using WordNet, API names, and crowd knowledge to reduce NL--code mismatch~\cite{lu2015query,nie2016query,lemos2014thesaurus,satter2016search,zhang2017expanding}. Neural methods predict keywords and control over-expansion~\cite{liu2019neural,huang2018query,huang2019deep}.

GACR~\cite{li2022generation} augments documentation queries with generated code snippets, substantially improving CodeSearchNet across languages. SSQR~\cite{mao2023self} treats reformulation as self-supervised masked completion (T5), improving MRR on CodeBERT/Lucene without parallel data. SG-BERT/GPT2~\cite{liu2023self} trains on large query corpora to generate expansions that lift top-$k$ accuracy. Hybrid embedding+sequence models (ECSQE) further enhance ranking~\cite{bibi2025enhancing}. Transformers and pretraining provide stronger semantics for bridging NL queries and code artifacts~\cite{vaswani2017attention}.
Practical note: many systems enforce symbol dictionary/namespace constraints and parser/compiler checks to avoid injecting invalid identifiers.

\subsection{Lessons Learned Across Applications}
Across domains, three recurring themes emerge. First, traditional QE remains valuable for efficiency and transparency, particularly when latency budgets are tight or supervision is scarce. Second, PLM/LLM-based QE often brings consistent gains in contextual disambiguation, cross-domain or cross-lingual adaptation, and recall--precision balance, especially when evidence signals are available to help control drift. Third, cost, controllability, and validity remain key deployment barriers; in practice, systems frequently rely on selective invocation, conservative fusion, ontology/KB integration, and lightweight validation to keep expansions aligned with domain constraints.

\section{Practical Deployment Considerations for Query Expansion}
\label{sec:blueprint}

The preceding sections reviewed query expansion (QE) methods by model family, point of injection, grounding and interaction pattern, learning/alignment strategy, and structured knowledge integration. Taken together, these sections suggest that deployment choices should not be framed as a simple contrast between traditional and LLM-based QE. Rather, the practical question is which QE family offers the best trade-off between controllability, evidence dependence, cost, and downstream risk in a given scenario. This section therefore distills deployment-oriented implications from the surveyed literature, without introducing a new taxonomy beyond the one already established in Secs.~\ref{subsec:Point_of_injection}--\ref{subsec:KG-augmented}.


\subsection{Deployment-Relevant Dimensions Emerging from the Survey}
\label{sec:blueprint_dimensions}

Across the surveyed work, four dimensions repeatedly shape how QE methods behave in deployment. These dimensions help explain why different QE variants fit different deployment constraints.

\paragraph{Where expansion is injected.}
A central distinction is where expansion enters the pipeline. Implicit embedding-based QE refines the query representation without emitting new terms, as in ANCE-PRF, ColBERT-PRF, Eclipse, and QB-PRF~\citep{yu2021improving,wang2023colbert,d2025eclipse,zhang2024selecting}. Explicit QE instead modifies the text side, either by selecting candidate terms or spans~\citep{naseri2021ceqe,naseri2022ceqe,zheng2020bert,Khader2023LearningTR,bassani2023personalized} or by generating pseudo-documents, multi-queries, or auxiliary passages~\citep{wang-etal-2023-query2doc,jagerman2023query,gao2023precise,mao2021generation,mackie2023generative,liu-zhang-2025-exp4fuse}. In deployment, implicit and selection-based variants are typically easier to constrain, while free-form generation is more prone to drifting off-topic or introducing unsupported content.

\paragraph{How strongly the expansion is grounded.}
Another axis is how the expansion is grounded. Some generative approaches rely mainly on model-internal knowledge and provide little explicit evidence for the added content~\citep{wang-etal-2023-query2doc,jagerman2023query,gao2023precise}. Other methods anchor expansion to retrieved evidence or structured sources within a single pass~\citep{jia2024mill,chen2024analyze,chuang2023expand,wang2023generative,lei2024corpus,zhang2024exploring,li2025pseudo,jaenich2025fair}. Interactive designs use intermediate retrieval results as feedback and revise expansion over multiple rounds~\citep{feng2024synergistic,rashid2024progressive,shen-etal-2024-retrieval,lei-etal-2025-thinkqe}. In practice, the key question is whether richer expansions remain tied to observable evidence, and whether that evidence basis can be refreshed during interaction.

\paragraph{When cost is paid.}
Methods also differ in when computation is incurred. Prompting and interactive generation increase serving-time work, affecting latency and cost per query. Alignment and distillation shift part of this cost offline by transferring useful expansion behavior into cheaper models or representations~\citep{pimpalkhute2024softqe,lee2024radcot,yao2025expandr,yang2025aligned}.

\paragraph{How structured external knowledge is used.}
A further dimension is the use of structured knowledge. Some approaches retrieve entities and relations directly from knowledge graphs, as in KGQE~\cite{perna2025knowledge} and CL-KGQE~\cite{rahman2019query}. Others build hybrid graphs that combine knowledge graph structure with document context, as in KAR~\cite{xia-etal-2025-knowledge} and QSKG~\cite{mackie2022query}. Both depend on entity linking quality and graph coverage. Direct KG retrieval tends to enforce stronger semantic consistency, while hybrid constructions often match collection-specific vocabulary better but can be more sensitive to linking and context-fusion noise.

 \subsection{Comparative Synthesis of QE Families for Deployment}
\label{sec:scorecard}

To make the practical implications of the surveyed literature easier to use in real systems,
Table~\ref{tab:qe_deployment_synthesis} summarizes the major QE families through a common
deployment lens. The descriptions synthesize recurring tendencies reported across representative
studies and are intended as qualitative deployment guidance rather than as universal rankings or
fixed prescriptions.

\begin{table*}[t]
\centering
\caption{Deployment-oriented comparison of QE families reviewed in this survey. The table summarizes typical use settings, grounding and control characteristics, major risks, operating burden, and practical guidance suggested by representative studies.}
\label{tab:qe_deployment_synthesis}
\footnotesize
\renewcommand{\arraystretch}{1.14}
\begin{tabularx}{\textwidth}{L{2.55cm} L{2.1cm} X X L{1.75cm} L{2.2cm}}
\toprule
\textbf{QE family} & \textbf{Typical use setting} & \textbf{Grounding / control} & \textbf{Main risks} & \textbf{Operating burden} & \textbf{Practical guidance} \\
\midrule

Traditional corpus-/resource-based QE~\citep{rocchio1971relevance,voorhees1994query,lavrenko2017relevance}
& Sparse retrieval; tight latency
& Grounded in corpus statistics or curated resources; relatively interpretable and controllable
& Depends on first-pass quality or static resources; may introduce noisy or overly generic terms under weak feedback
& Low
& Strong baseline under efficiency, stability, and interpretability constraints \\
\addlinespace[2pt]

Implicit embedding-based QE~\citep{yu2021improving,wang2023colbert,d2025eclipse,zhang2024selecting}
& Dense or late-interaction retrieval
& No explicit added terms; usually conservative at the surface-form level; limited pipeline changes
& Still depends on first-pass retrieval or feedback quality; gains may weaken when initial retrieval is poor
& Low to medium
& Suitable for low-risk refinement in neural retrieval pipelines \\
\addlinespace[2pt]

Selection-based explicit QE~\citep{naseri2021ceqe,naseri2022ceqe,zheng2020bert,Khader2023LearningTR,bassani2023personalized}
& Explicit and controllable expansion
& Bounded by candidate pools or selection rules; produces interpretable terms or spans
& Limited by candidate coverage; cannot introduce concepts outside the available pool
& Low to medium
& Preferable when bounded behavior, auditability, and control matter more than flexibility \\
\midrule

Zero-grounding single-round generative QE~\citep{wang-etal-2023-query2doc,jagerman2023query,mao2021generation,mackie2023generative,gao2023precise,liu-zhang-2025-exp4fuse}
& Zero-shot or weak-resource settings
& Mainly grounded in model-internal knowledge and prompting; weaker external control
& Typically more exposed to unsupported additions, topic drift, and model/version sensitivity
& Medium to high
& Best used selectively for hard or underspecified queries, rather than as an always-on default \\
\addlinespace[2pt]

Retrieval-grounded single-round generative QE~\citep{jia2024mill,chen2024analyze,chuang2023expand,wang2023generative,lei2024corpus,zhang2024exploring,li2025pseudo,jaenich2025fair}
& RAG, domain retrieval, evidence-sensitive search
& Expansion is anchored in retrieved passages or pseudo-relevant evidence; typically offers stronger grounding than unguided generation
& Inherits errors, omissions, and bias from first-pass evidence; effectiveness depends on grounding depth and filtering
& Medium to high
& Often preferable when faithfulness matters and an initial evidence pool is available \\
\addlinespace[2pt]

Interactive multi-round QE~\citep{feng2024synergistic,rashid2024progressive,shen-etal-2024-retrieval,lei-etal-2025-thinkqe}
& Ambiguous, complex, or hard queries
& Iteratively refines expansion using refreshed evidence or interaction; offers stronger adaptive control
& Highest latency and engineering burden; errors may accumulate across rounds if the loop is not well controlled
& High
& Better suited to difficult or high-value queries than to universal default deployment \\
\addlinespace[2pt]

KG-augmented QE~\citep{perna2025knowledge,rahman2019query,xia-etal-2025-knowledge,mackie2022query}
& Entity-centric or semi-structured retrieval
& Strong entity- and relation-level control when structured knowledge is reliable
& Depends on entity linking, graph coverage, and freshness; may add integration overhead
& Low to medium
& Useful as a complement when structured knowledge is central and graph quality is reliable \\
\bottomrule
\end{tabularx}
\end{table*}

Several deployment implications follow from Table~\ref{tab:qe_deployment_synthesis}.
When latency, stability, and predictable behavior are dominant constraints, the literature reviewed here often points to traditional QE, implicit embedding-based QE, or selection-based explicit QE as conservative starting points~\citep{yu2021improving,wang2023colbert,naseri2021ceqe,zheng2020bert}.
When the downstream task is evidence-sensitive, as in RAG or domain-specific QA, the surveyed work more often couples expansion with retrieved passages, structured resources, or explicit filtering rather than relying on purely free-form generation~\citep{jia2024mill,chen2024analyze,perna2025knowledge,xia-etal-2025-knowledge}.
Interactive multi-round methods are promising for ambiguous or difficult queries, but the studies reviewed in Sec.~\ref{subsec:grounding_interaction} suggest that they are best viewed as expensive adaptive retrieval strategies rather than universal defaults~\citep{feng2024synergistic,rashid2024progressive,shen-etal-2024-retrieval,lei-etal-2025-thinkqe}.
Finally, supervision, preference alignment, and distillation are better understood as cross-cutting optimization strategies rather than standalone QE families, especially when repeated traffic justifies offline investment~\citep{pimpalkhute2024softqe,lee2024radcot,yao2025expandr,yang2025aligned}.

\subsection{Evaluation and Reporting for Deployment}
\label{sec:eval_analysis}

Having outlined a deployment-oriented blueprint for selecting and combining QE strategies, we finally turn to how such choices should be evaluated and reported in practice. QE evaluation is most informative when it is aligned with the role QE plays in the deployed pipeline, rather than reduced to a single end-task gain. In practice, QE is judged not only by retrieval effectiveness, but also by controllability (whether expansions follow intended constraints), grounding (whether retrieved evidence supports downstream outputs), and operating cost.

The relative importance of evaluation metrics depends on what QE is expected to improve. Traditional QE is often assessed with an emphasis on coverage and deeper recall, reflecting its role in alleviating vocabulary mismatch and recovering otherwise-missed relevant documents~\citep{carpineto2012survey,azad2019query}. In evidence-sensitive pipelines such as RAG or open-domain QA, however, expansions that introduce off-topic or unverifiable content may contaminate the evidence set and propagate errors downstream. In such settings, early precision and evidence support (e.g., answer--evidence consistency) can be as important as deep recall, depending on the downstream verification and synthesis stage~\citep{zhang2025siren,niu2024ragtruth,koo2024optimizing,huquery}. Top-heavy metrics such as MRR@10 and nDCG@10 mainly reward improvements in early precision, whereas Recall@1000 and related coverage-oriented measures are more informative when QE is intended to recover missed relevant documents. MAP and AP remain useful for connecting newer methods back to the classical ad hoc IR literature, although they are now reported less consistently in PLM/LLM-based work.

For PLM/LLM-based QE, method specification and cost should be treated as part of the evaluation protocol. Methods such as Query2Doc, CoT-QE, HyDE, MILL, AGR, and RADCoT differ not only in effectiveness, but also in prompt design, evidence construction, decoding strategy, the number of retrieval passes, and whether expensive generation is reduced through distillation~\citep{wang-etal-2023-query2doc,jagerman2023query,gao2023precise,jia2024mill,chen2024analyze,lee2024radcot}. Deployment-oriented reporting therefore benefits from documenting the model or API version, decoding settings, prompts or templates, pseudo-evidence construction, the number of LLM calls and retrieval passes, and token usage or latency when these materially affect cost.

Robustness analysis is most diagnostic when tied to application-specific failure modes. Conversational search is sensitive to stale history and context selection~\citep{mo2023learning,kumar2020making,voskarides2020query}; code search requires symbol validity and repository-specific grounding~\citep{li2022generation,mao2023self,liu2023self}; KG-augmented or domain-specific retrieval depends on entity linking quality and structured evidence coverage~\citep{perna2025knowledge,wu2025kg,xia-etal-2025-knowledge}. Under such conditions, aggregate benchmark scores can obscure practically important differences, making it more informative to stratify results by query type, ambiguity, domain specificity, or downstream use case.

Benchmark choice also introduces systematic differences in what is being measured.
MS MARCO is derived from real anonymized user queries and has become a central benchmark for large-scale passage or document retrieval and ranking~\citep{DBLP:conf/nips/NguyenRSGTMD16}.
As a result, it is especially useful for assessing whether QE improves ranking effectiveness in realistic web-style retrieval conditions.
TREC Deep Learning, by contrast, focuses on passage and document ranking with smaller but more deeply judged test sets and graded relevance labels, making it useful for more reliable top-ranked evaluation~\citep{craswell2025overview}.
BEIR is valuable for testing domain transfer and zero-shot robustness because it spans diverse retrieval tasks and domains, but its heterogeneity also means that aggregate averages may obscure where a QE method actually helps or fails~\citep{thakur2021beir}.
Accordingly, benchmark results should be interpreted with respect to task type, judgment depth, and the intended deployment setting rather than treated as universally interchangeable.

\section{Future Directions and Open Challenges}
\label{sec:future}

Pre-trained and large language models have substantially advanced query expansion (QE), but several challenges remain unresolved before these methods can be deployed reliably at scale. Two issues are especially important. First, expansions must be accurate, inspectable, and safe, particularly in evidence-sensitive or high-stakes applications where unsupported additions can harm retrieval quality and propagate errors downstream. Second, QE systems must remain efficient and adaptable as queries, corpora, and external knowledge evolve over time. This section discusses these open challenges and outlines promising directions for improving quality control, safety, efficiency, and continual adaptation.

\subsection{Quality control and safety
}
\label{sec:future_safety}

Even strong QE methods can reduce precision when the generated or selected expansion terms drift away from the original query intent, and automated filtering remains imperfect under noisy, ambiguous, or adversarial conditions~\cite{weller2024generative}. In high-stakes settings, introducing unsupported entities, relations, or facets may pollute the retrieved evidence set and mislead downstream ranking, reading, or generation components.

Different categories of QE methods mitigate drift in different ways, but none eliminates the problem entirely. Prompt-based and generative methods may introduce fluent but weakly grounded content; retrieval-grounded methods depend on the quality of the evidence source; and structured or domain-specific methods can still fail when entity linking, schema alignment, or coverage is incomplete. 

Several research directions appear particularly important. First, the field needs low-cost faithfulness checks that go beyond expensive LLM-as-a-judge pipelines. Promising directions include lightweight consistency constraints, lexical or semantic evidence-overlap tests, and selective verification that is triggered only for uncertain cases. Second, future QE systems should make expansions more auditable by attaching provenance whenever possible, such as supporting evidence snippets, retrieved passages, or KG facts for newly introduced entities or facets. This would make expansions easier to inspect, justify, and roll back when errors are detected. 
Third, LLM-based QE faces a practical tension between informativeness and consumability. On the one hand, expansions should introduce diverse yet relevant cues to mitigate underspecification and vocabulary mismatch; on the other hand, they should remain coherent and well-formed so that downstream components (e.g., LLM rerankers or RAG readers) can reliably interpret and use them.
This motivates explicit controls on expansion style (keyword-rich vs.\ natural language).
Fourth, more work is needed on monotonicity-oriented testing, where a candidate expansion is accepted only if it does not harm retrieval under controlled counterfactual checks, such as drop-term, swap-sense, or perturbation-based comparisons.

\subsection{Efficiency and continual adaptation}

Always-on LLM-based QE remains costly, especially when expansion requires multiple prompts, retrieval passes, or large-model calls. Techniques such as selective invocation, semantic caching, distillation \cite{yang2025survey}, and parameter-efficient adaptation are therefore essential for deployment, yet many existing studies still under-report operating cost, latency, and freshness-related trade-offs.

Another open issue is temporal adaptation \cite{wang2024large}. Static parametric models cannot keep pace with emerging events, new terminology, or domain shifts, which means that purely model-internal expansions may become stale. From a deployment perspective, QE should therefore be treated not as a one-time component, but as a lifecycle process in which expansions are generated, monitored, refreshed, and, when necessary, rolled back.

Several directions are especially promising here. First, freshness-aware caching deserves more attention: deployable systems should explicitly model when cached expansions remain useful, when they should be refreshed, and how stale outputs can be detected before they harm retrieval quality. Second, distillation and lightweight adaptation methods should be designed to preserve not only efficiency, but also expansion diversity and cross-domain robustness; otherwise, smaller deployed models may collapse to narrow or overly conservative expansion behavior. 
Third, future work can move toward version-controlled QE deployment, where prompt templates, model versions, evidence sources, and decoding settings are recorded systematically so that updates can be evaluated through regression tests.
Such practices would make QE pipelines easier to compare, maintain, and roll back in real-world settings.

\section{Conclusion}
\label{sec:conclusion}

This survey traces the evolution of query expansion (QE) from classical corpus- and lexicon-based methods to contemporary PLM/LLM paradigms. We organize the literature along four complementary dimensions, namely point of injection, grounding and interaction, learning and alignment, and knowledge-graph integration, to clarify how QE methods differ in their mechanisms, assumptions, and trade-offs across retrieval pipelines. We also briefly discuss practical deployment considerations and summarize open challenges and future directions related to quality control, safety, efficiency, and continual adaptation. Overall, we hope this survey provides a structured reference for understanding, comparing, and applying QE methods in modern retrieval systems.

\bibliographystyle{ACM-Reference-Format}
\bibliography{reference}

\end{document}